\documentclass[journal=jacsat,manuscript=article]{achemso}
\usepackage{comment}

\usepackage[T1]{fontenc} 
\usepackage{amsmath} 
\usepackage{multirow}
\usepackage{tabularx}
\usepackage{booktabs}
\usepackage{adjustbox}
\usepackage{xcolor}

\newcommand*\mycommand[1]{\texttt{\emph{#1}}}

  \title[An \textsf{achemso} demo]{ALD-Derived WO\textsubscript{3-x} Leads to Nearly Wake-Up-Free Ferroelectric Hf\textsubscript{0.5}Zr\textsubscript{0.5}O\textsubscript{2} at Elevated Temperatures}

\author{Nashrah Afroze}
\affiliation[GT]{Department of Electrical and Computer Engineering, Georgia Institute of Technology, Atlanta, GA-30332, USA.}
\email{nafroze3@gatech.edu}
\altaffiliation{These authors contributed equally to this work.}

\author{Jihoon Choi}
\affiliation[UNIST]{School of Energy and Chemical Engineering, Ulsan National Institute of Science and Technology (UNIST), Ulsan-44919, South Korea.}
\email{jihoonchoi@unist.ac.kr}
\altaffiliation{These authors contributed equally to this work.}

\author{Salma Soliman}
\affiliation[GT]{Department of Electrical and Computer Engineering, Georgia Institute of Technology, Atlanta, GA-30332, USA.}

\author{Chang Hoon Kim}
\affiliation[UNIST]{School of Energy and Chemical Engineering, Ulsan National Institute of Science and Technology (UNIST), Ulsan-44919, South Korea.}

\author{Jiayi Chen}
\affiliation[GT]{Department of Electrical and Computer Engineering, Georgia Institute of Technology, Atlanta, GA-30332, USA.}

\author{Yu-Hsin Kuo}
\affiliation[GT]{Department of Electrical and Computer Engineering, Georgia Institute of Technology, Atlanta, GA-30332, USA.}

\author{Mengkun Tian}
\affiliation[GT]{Institute of Materials and Systems, Georgia Institute of Technology, GA-30332, USA.}

\author{Chengyang Zhang}
\affiliation[GT]{Department of Electrical and Computer Engineering, Georgia Institute of Technology, Atlanta, GA-30332, USA.}

\author{Priyankka Gundlapudi Ravikumar}
\affiliation[GT]{Department of Electrical and Computer Engineering, Georgia Institute of Technology, Atlanta, GA-30332, USA.}

\author{Suman Datta}
\affiliation[GT]{Department of Electrical and Computer Engineering, Georgia Institute of Technology, Atlanta, GA-30332, USA.}
\alsoaffiliation[GT2]{Department of  Materials Science and Engineering, Georgia Institute of Technology, Atlanta, GA-30332, USA.}

\author{Andrea Padovani}
    \affiliation[UNIMORE]{Department of Engineering Sciences and Methods (DISMI), University of Modena and Reggio Emilia, 42122 Reggio Emilia, Italy.}

\author{Jun Hee Lee}
\affiliation[UNIST]{School of Energy and Chemical Engineering, Ulsan National Institute of Science and Technology (UNIST), Ulsan-44919, South Korea.}
\alsoaffiliation[UNIST2]{Graduate School of Semiconductor Materials and Devices Engineering, Ulsan National Institute of Science and Technology (UNIST), Ulsan-44919, South Korea.}
\email{junhee@unist.ac.kr}

\author{Asif Khan}
\affiliation[GT]{Department of Electrical and Computer Engineering, Georgia Institute of Technology, Atlanta, GA-30332, USA.}
\alsoaffiliation[GT2]{Department of  Materials Science and Engineering, Georgia Institute of Technology, Atlanta, GA-30332, USA.}
\email{akhan40@gatech.edu}

\abbreviations{IR,NMR,UV}
\keywords{American Chemical Society, \LaTeX}
\begin{document}

\begin{abstract}
Breaking the memory wall in advanced computing architectures will require complex 3D integration of emerging memory materials such as ferroelectrics—either within the back-end-of-line (BEOL) of CMOS front-end processes or through advanced 3D packaging technologies. Achieving this integration demands that memory materials exhibit high thermal resilience, with the capability to operate reliably at elevated temperatures such as 125 $^\circ$C, due to the substantial heat generated by front-end transistors. However, silicon-compatible HfO\textsubscript{2}-based ferroelectrics tend to exhibit antiferroelectric-like behavior in this temperature range, accompanied by a more pronounced wake-up effect, posing significant challenges to their thermal reliability. Here, we report that by introducing a thin tungsten oxide (WO\textsubscript{3-x}) layer—known as an oxygen reservoir—and carefully tuning its oxygen content, ultra-thin Hf\textsubscript{0.5}Zr\textsubscript{0.5}O\textsubscript{2} (5 nm) films can be made robust against the ferroelectric-to-antiferroelectric transition at elevated temperatures. This approach not only minimizes polarization loss in the pristine state but also effectively suppresses the wake-up effect, reducing the required wake-up cycles from 10\textsuperscript{5} to only 10 at 125 $^\circ$C- a qualifying temperature for back-end memory integrated with front-end logic, as defined by the JEDEC standard. First-principles density functional theory (DFT) calculations reveal that WO\textsubscript{3} enhances the stability of the ferroelectric orthorhombic phase (o-phase) at elevated temperatures by increasing the tetragonal-to-orthorhombic phase energy gap, and promoting favorable phonon mode evolution, thereby supporting o-phase formation under both thermodynamic and kinetic constraints.

\end{abstract}

\noindent\textbf{Keywords:} Ferroelectric memory, HfO$_2$-based ferroelectrics, Oxygen reservoir layer, ALD deposited WO$_{3-x}$, High-temperature operation, Nearly wake-up-free, Memory-on-logic, 3D integration.

\section{Introduction}

AI is fueling advances across domains such as high-performance computing, cloud infrastructure, mobile platforms, autonomous vehicles, and augmented reality, yet the massive datasets and complex models driving these applications are turning training and inference into critical memory and reliability challenges. Ferroelectric memory technologies, such as ferroelectric NAND (FE-NAND), ferroelectric random access memory (Fe-RAM), and ferroelectric field effect transistors (FeFETs) have become leading contenders for nonvolatile memory solutions across various segments of the memory hierarchy\cite{Khan2020,mikolajick2018ferroelectric,10413697}. Recently, significant advancements in ferroelectric memories have been achieved, including the demonstration of a high-capacity dual-layer FE 1T-1C memory chip with a 32 Gb capacity and DRAM-comparable performance, as well as emerging three-dimensional integration of ferroelectric devices, exemplified by vertically stacked multi-layer ferroelectric architectures for high-density integration\cite{ramaswamy2023nvdram,Huang2025}. Despite advantages such as non-volatility and high charge density, ferroelectric materials require a deeper understanding to ensure reliable operation, particularly under elevated temperatures \cite{chavan2024materials}. 

As the semiconductor industry advances toward 3D-IC integration, the intermediate step of memory-on-logic has demonstrated benefits comparable to a full technology node improvement. This memory-on-logic configuration shortens the communication distance between logic and memory, delivering up to 22\% higher performance and 36\% lower power consumption \cite{9502475}. However, thermal management poses a critical challenge. With an increasing number of stacked dies, tiers located farther from the heat sink and closer to heat-generating logic layers experience substantial thermal buildup due to thermal resistance and crosstalk \cite{zhang2014thermal} (Figure 1a). Since logic dies are generally capable to operate at 105-125\textdegree C\cite{zhou2019temperature}, memories must also be qualified to operate reliably at 125\textdegree C according to the JEDEC JESD22-A108 standard for 3D-IC qualification. Figure 1b summarizes the maximum operating temperature specifications of Micron DRAMs across different application scenarios\cite{Micron_LPDDR5_2023}. These specifications highlight the need for robust high-temperature performance, which is essential for enabling reliable integration of ferroelectric memories into emerging heterogeneous and monolithic 3D (H3D and M3D) systems and memory-on-logic architectures. Unfortunately, HfO\textsubscript{2}-based ferroelectrics, which form the foundation of many next-generation memory devices, remain highly vulnerable to performance degradation at elevated temperatures. It is well-established that increasing temperature induces an orthorhombic-to-tetragonal phase transition in ferroelectric materials, leading to pinching of the polarization loops and a gradual decrease in remnant polarization\cite{boscke2011phase,mehmood2020wake,park2018origin,mimura2021large,schroeder2022temperature,mittmann2020impact,hoffmann2015ferroelectric}. Consequently, ferroelectricity can be lost under thermal stress, severely limiting device reliability\cite{10019561,mittmann2020impact,sunbul2023impact,ravikumar2024comprehensive}. Although their has been lots of studies done on improving wakeup behavior and orthorhombic phase enhancement at room temperature using strategies like different atomic layer deposition (ALD) techniques \cite{doi:10.1021/acsami.4c18056}, electrode engineering\cite{chen2024uniform, alcala2023electrode, Wang2023, wang2024modulation}, controlling Oxygen flow during deposition \cite{doi:10.1021/acsaelm.0c00671},high temperature cycling\cite{feng2025record} and interface engineering \cite{ yang2023wake, lee2022modulating, Shi2023}, systematic efforts to improve wakeup effect and pristine state polarization at elevated temperatures remain scarce in the literature\cite{lehninger2023ferroelectric}.

Among various strategies to enhance ferroelectric device performance, interface engineering has shown particular promise. Several interfacial layers have been explored to improve endurance, leakage, and polarization characteristics. For example, NbO\textsubscript{2} and TiO\textsubscript{2} have been used for endurance enhancement\cite{popovici2022high,itoya2024enhanced,doi:10.1021/acsomega.2c06237}, Pt has been employed to reduce leakage and wake-up effects\cite{doi:10.1021/acsaelm.0c00671}, and TiON and WN\textsubscript{x} have been investigated for boosting polarization\cite{lee2022modulating,chen2023controlling,aich2024low}. More recently, WS\textsubscript{2} has demonstrated improvements in both polarization and endurance\cite{10873482}. In particular, WO\textsubscript{x} has gained attention as an effective oxygen reservoir for improving different properties of ferroelectric capacitors at room temperature\cite{yang2023wake,kwon2024improvement,choi2024oxygen,doi:10.1021/acsaelm.3c01502,kim2025comprehensive,doi:10.1021/acsami.4c18056, zhao2023engineering,zhao2025ultrathin, wang2024optimal, habibi2025highly}. However, its potential for enhancing ferroelectric performance at elevated temperatures remains largely unexplored\cite{afroze2024self,10983715,zhang2025enhanced}.

\begin{figure}[H]
\centering
\includegraphics[ width=6.5in]{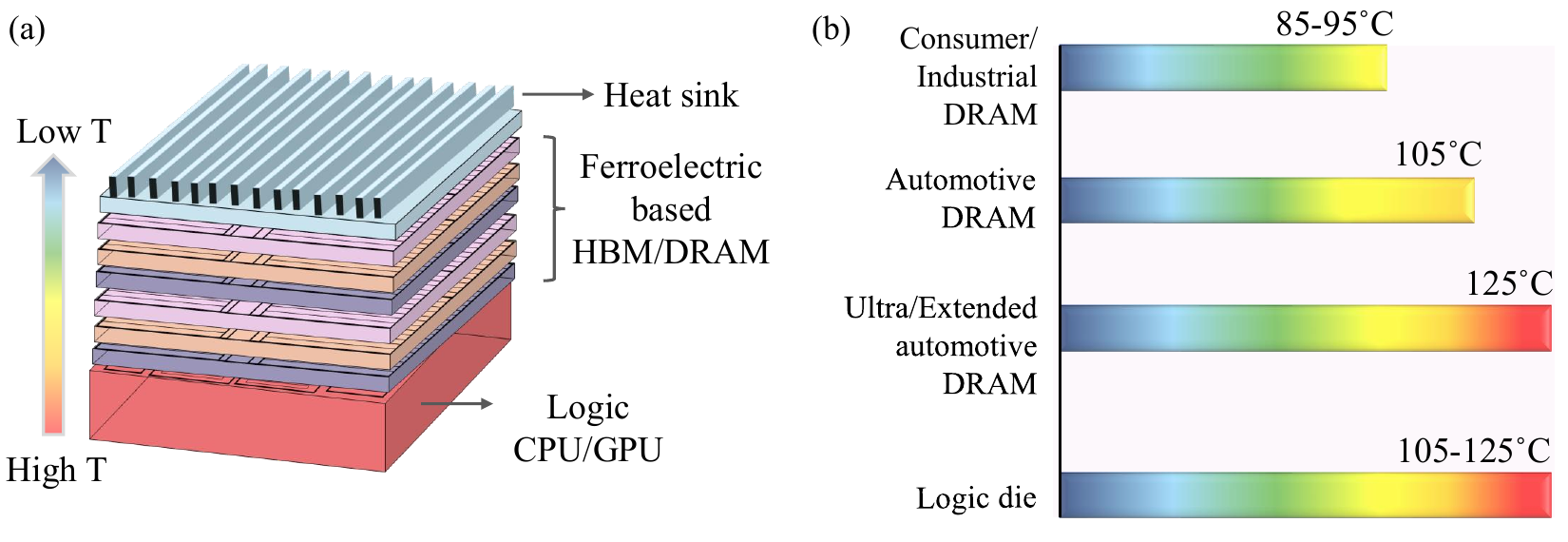}
\caption{Importance of high temperature performance enhancement of ferroelectric memories. (a) Schematic of 3D integration of memory on logic architecture. Memories away from heat sink gets heated due to the heated logic die. (b) Operating temperatures of Micron DRAMs based on different applications.   
 }
\end{figure}

In this study, we demonstrate that incorporating WO\textsubscript{3-x} as an interfacial oxygen reservoir layer significantly enhances the high-temperature performance of Hf\textsubscript{0.5}Zr\textsubscript{0.5}O (HZO)-based capacitors. WO\textsubscript{3-x} is introduced at the interface between the bottom electrode and the HZO film, either by oxidizing the electrode via O\textsubscript{2} plasma treatment or through atomic layer deposition (ALD). This interface engineering approach increases the ability to retain orthorhombic phase in HZO, suppressing the emergence of antiferroelectric behavior in pristine devices under thermal stress. As a result, the number of bipolar cycles required to achieve wake-up is significantly reduced in WO\textsubscript{3-x}-incorporated devices. First-principles density functional theory (DFT) calculations indicate that the presence of WO\textsubscript{3} stabilizes the ferroelectric o- phase at elevated temperatures through a combination of effects: an increased tetragonal–orthorhombic phase energy gap in the Helmholtz free energy landscape and enhanced $X\textsubscript{2}'$ phonon mode coupling. These effects arise from the lower entropy and lattice anisotropy induced by WO\textsubscript{3}, which together favor orthorhombic Pca2\textsubscript{1} phase formation under both thermodynamic and kinetic considerations. 
This stabilization of ferroelectric behavior under thermal stress enhances the temperature resilience of the device. Our results provide a promising pathway for achieving thermally robust ferroelectric memories, such as Fe-RAMs and Fe-FETs, enabling their deployment in emerging 3D memory-on-logic architectures where elevated temperatures present a significant reliability challenge.

\section{Results and discussion}

\subsection{Experiment}

Figure 2a illustrates the fabrication process flow of ferroelectric capacitors incorporating WO\textsubscript{3-x} as an oxygen reservoir layer, introduced via two distinct methods. In the first approach, partial oxidation of the sputtered W bottom electrode on Si was carried out using O\textsubscript{2} plasma treatment within the atomic layer deposition (ALD) chamber, resulting in the formation of an approximately 4 nm thick WO\textsubscript{3-x} layer at the W surface. In the second approach, a plasma-enhanced ALD (PE-ALD) technique was employed to directly deposit 5 nm and 6 nm thick WO\textsubscript{3-x} layers on the W electrode.
For the reference device, the HZO layer was deposited directly on the bottom W electrode without any intermediate WO\textsubscript{3-x} layer. Further details of the fabrication procedure are described in the \textit{Methods} section.

 Scanning transmission electron microscopy (STEM) images of the cross-section of the fabricated devices are presented in the left most panel of Figures 2b–d, corresponding to the reference, O\textsubscript{2} plasma, and ALD-based WO\textsubscript{3-x} device structures, respectively. To verify the formation of WO\textsubscript{3-x} at the bottom interface in both the O\textsubscript{2} plasma and ALD-based devices, energy-dispersive X-ray spectroscopy (EDS) was carried out. Elemental mapping confirmed the presence of distinct W and O signals at the interface in these devices, validating the successful formation of a WO\textsubscript{3-x} layer. In contrast, the reference device exhibited no such signals, indicating the absence of an interfacial WO\textsubscript{3-x} layer.

\begin{figure}[H]
\centering
\includegraphics[ width=6.5in]{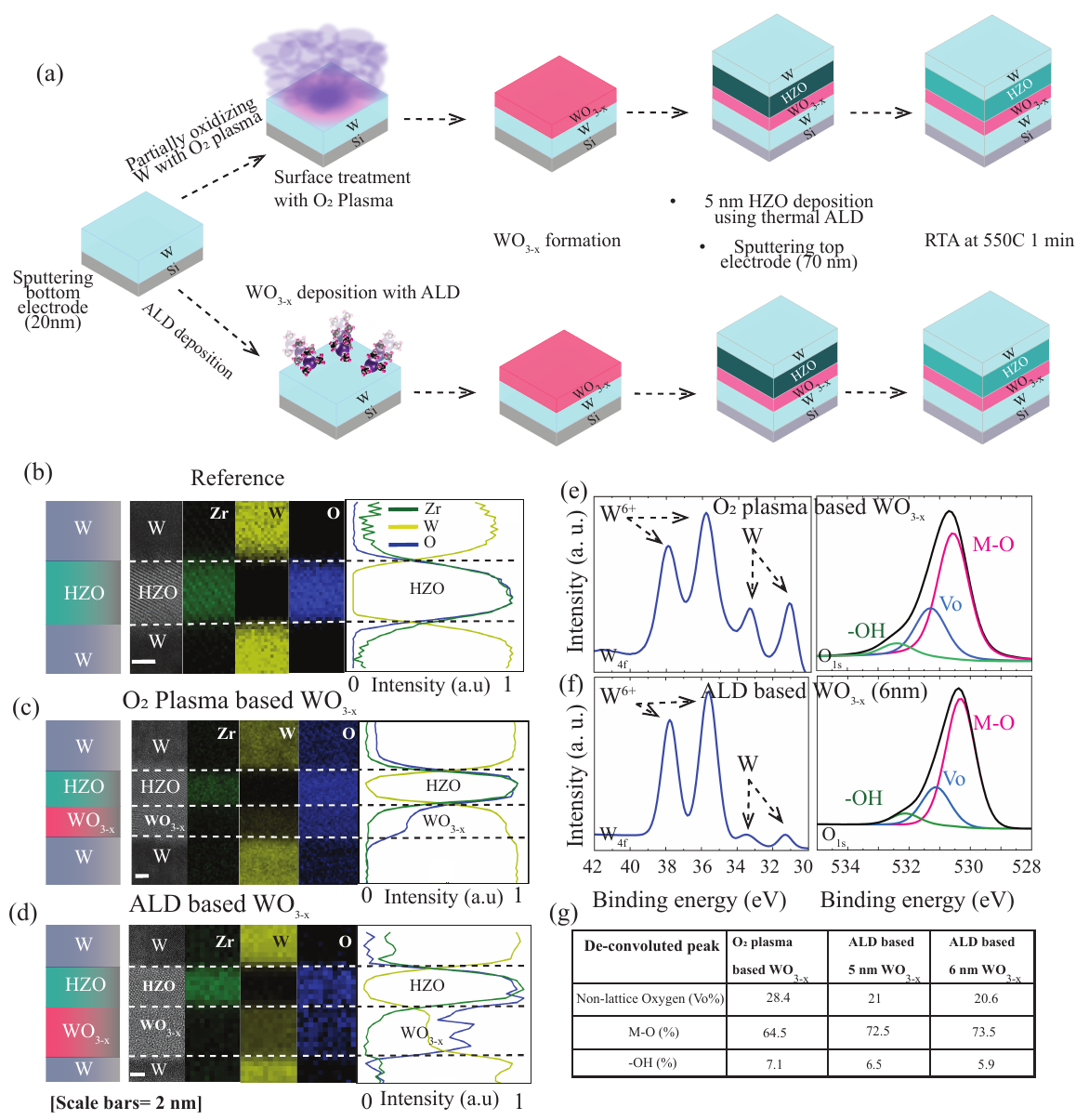}
\caption{Device structures and material characterization of ferroelectric capacitors. (a) Fabrication process of O\textsubscript{2} plasma and ALD based WO\textsubscript{3-x} devices. (b-d) Cross-sectional STEM images of (b) reference, (c) O\textsubscript{2} plasma and (d) ALD based 6 nm WO\textsubscript{3-x} devices along with their EDS characterization. Material-count map coming from Zr (green), W (yellow) and O (blue) and the corresponding line scans for (b) reference, (c) O\textsubscript{2} Plasma and (d) ALD devices. The line scans on the right of each material count map further confirms the layers and their interfaces. XPS spectra obtained from W 4f and O 1s orbitals of WO\textsubscript{3-x} from (e) O\textsubscript{2} plasma and , (f) ALD based 6nm WO\textsubscript{3-x} samples. (e,f) Magenta, blue and green curves in the O 1s plot are the de-convoluted peaks corresponding to M-O, non-lattice Oxygen (V\textsubscript{o}) and -OH respectively. (g) Table showing the percentage obtained from the deconvoluted peaks of O 1s scan of different samples.  
 }
\end{figure} 

To assess the stoichiometry of WO\textsubscript{3-x} films synthesized via O\textsubscript{2} plasma treatment and atomic layer deposition (ALD), X-ray photoelectron spectroscopy (XPS) analysis was conducted, as shown in Figures 2e and 2f. The W 4f spectra from both samples exhibit prominent W\textsuperscript{6+} peaks, confirming the successful formation of WO\textsubscript{3-x}. The O 1s spectra were deconvoluted to quantify the contributions from lattice and non-lattice oxygen species, with the extracted component ratios summarized in Figure 2g. Notably, the O\textsubscript{2} plasma based WO\textsubscript{3-x} film shows a significant contribution from non-lattice oxygen, corresponding to an oxygen vacancy (V\textsubscript{o}) concentration of approximately 28.4\%. In contrast, the ALD-deposited WO\textsubscript{3-x} films with 5 nm and 6 nm thicknesses exhibit a comparatively lower V\textsubscript{o} concentration of ~21\%. The full XPS spectrum for the 5 nm ALD-grown WO\textsubscript{3-x} is provided in Figure S1. These results indicate that while both methods yield substoichiometric WO\textsubscript{3-x}, the O\textsubscript{2} plasma process introduces a higher density of oxygen vacancies. Furthermore, these findings underscore that the deposition technique, rather than film thickness, plays a critical role in governing WO\textsubscript{3-x} stoichiometry, which in turn significantly impacts the electrical behavior of the devices, as elaborated in Figure 3.

Figure 3 presents the polarization–voltage (P–V) characteristics of all fabricated devices, along with the extracted coercive voltage and remnant polarization values from P-V and positive-up negative-down (PUND) measurements respectively. To decouple ferroelectric switching from leakage effects, PUND measurements were performed. All devices demonstrated comparable remnant polarization (2P\textsubscript{r}), confirming that the incorporation of WO\textsubscript{3-x} does not compromise ferroelectric polarization. Additionally, WO\textsubscript{3-x} plays a beneficial role in enhancing device endurance, both at room and at higher temperatures, as shown in Figure S2 and supported by previous studies \cite{kim2025comprehensive, choi2024oxygen}. A detailed analysis of temperature-dependent endurance improvements due to WO\textsubscript{3-x} incorporation can be found in earlier work \cite{afroze2024self}. Moreover, the inclusion of WO\textsubscript{3-x} does not adversely affect imprint and retention behavior, as illustrated in Figure S3 and S4. 

The coercive voltage of the O\textsubscript{2} plasma based device closely matches that of the reference device, whereas the ALD-deposited WO\textsubscript{3-x} devices exhibit higher coercive voltage (Figure 3). This behavior indicates that the ALD-deposited WO\textsubscript{3-x} is less conductive compared to the plasma based counterpart, thereby dropping more voltage across the WO\textsubscript{3-x} layer and requiring a higher switching voltage for the overlying HZO layer. This increase in coercive voltage is consistent with the comparatively lower conductivity measured on these devices (Figure S5), and the lower oxygen vacancy (V\textsubscript{o}) concentration observed in the ALD samples from the XPS analysis in Figures 2e–g. Given that oxygen vacancies significantly affect the electrical conductivity of WO\textsubscript{3} films \cite{nano11082136, habibi2025highly}, the reduced V\textsubscript{o} content in ALD-deposited layers leads to decreased conductivity.

\begin{figure}[H]
\centering
\includegraphics[ width=2.5in]{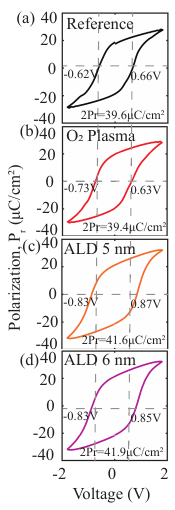}
\caption{Electrical characterization at room temperature. PV loops after 10\textsuperscript{5} cycles from (a) reference, (b) O\textsubscript{2} plasma-, ALD based (c) 5 nm and (d) 6 nm WO\textsubscript{3-x} devices. 2P\textsubscript{r} values are extracted from PUND measurement.   }
\end{figure}

The polarization–voltage (P–V) and switching current–voltage (I\textsubscript{SW}–V) characteristics of all three devices in their pristine states were measured across a wide range of temperatures, as shown in Figure 4. At room temperature, both the reference and the O\textsubscript{2} plasma based WO\textsubscript{3-x} devices exhibit similar P–V and I\textsubscript{SW}–V responses, while the ALD-deposited WO\textsubscript{3-x} device shows marginally enhanced ferroelectric behavior (Figures 4a–c). Upon increasing the temperature to 85\textdegree C, the reference device that lacks any WO\textsubscript{3-x} interfacial layer exhibits a pronounced antiferroelectric-like response, evidenced by a well-defined double-peak structure in the I\textsubscript{SW}–V curve (Figure 4d),indicating that the transition away from the orthorhombic ferroelectric phase initiates well before the targeted operating temperature (125\textdegree C) is reached.

In contrast, the O\textsubscript{2} plasma based device demonstrates improved thermal stability of ferroelectricity, evidenced by reduced separation of switching current peaks compared to reference device at 85\textdegree C (Figure 4e). A smaller separation between same-direction switching peaks indicates reduced antiferroelectric-like behavior\cite{xu2024doped}. Further enhancement in thermal robustness is observed for devices with ALD based WO\textsubscript{3-x} for the 6 nm layer (Figure 4f) and similarly for the 5 nm variant (Figure S6).  These devices show a minor residual switching peak on the negative voltage side across all temperatures, suggesting that WO\textsubscript{3-x} incorporation effectively suppresses or delays the onset of the antiferroelectric phase transition.

This trend is validated by PUND measurements performed under varying temperature conditions at pristine states. Using 200 ns square pulses of varying amplitudes, the switched polarization (2P\textsubscript{SW}) was extracted. At room temperature, all devices exhibit comparable saturated polarization values in pristine condition (Figure 4g). However, at 125\textdegree C, the reference device shows a significant drop in polarization relative to the WO\textsubscript{3-x}-containing devices (Figure 4h), indicating temperature-induced degradation of ferroelectricity in the absence of WO\textsubscript{3-x}. This observation is further reinforced by long-pulse (10 $\mu$s/2V) PUND measurements across a wide temperature range on pristine devices, where the extracted 2P\textsubscript{r} values confirm that ferroelectric property deteriorates significantly in the reference device, while devices with WO\textsubscript{3-x} maintain polarization more effectively (Figure 4i).These results highlight the crucial role of WO\textsubscript{3-x} in stabilizing the ferroelectric orthorhombic phase at elevated temperatures at pristine state, thereby preserving ferroelectric properties and suppressing unwanted phase transitions.

\begin{figure}[H]
\centering
\includegraphics[ width=6.5 in]
{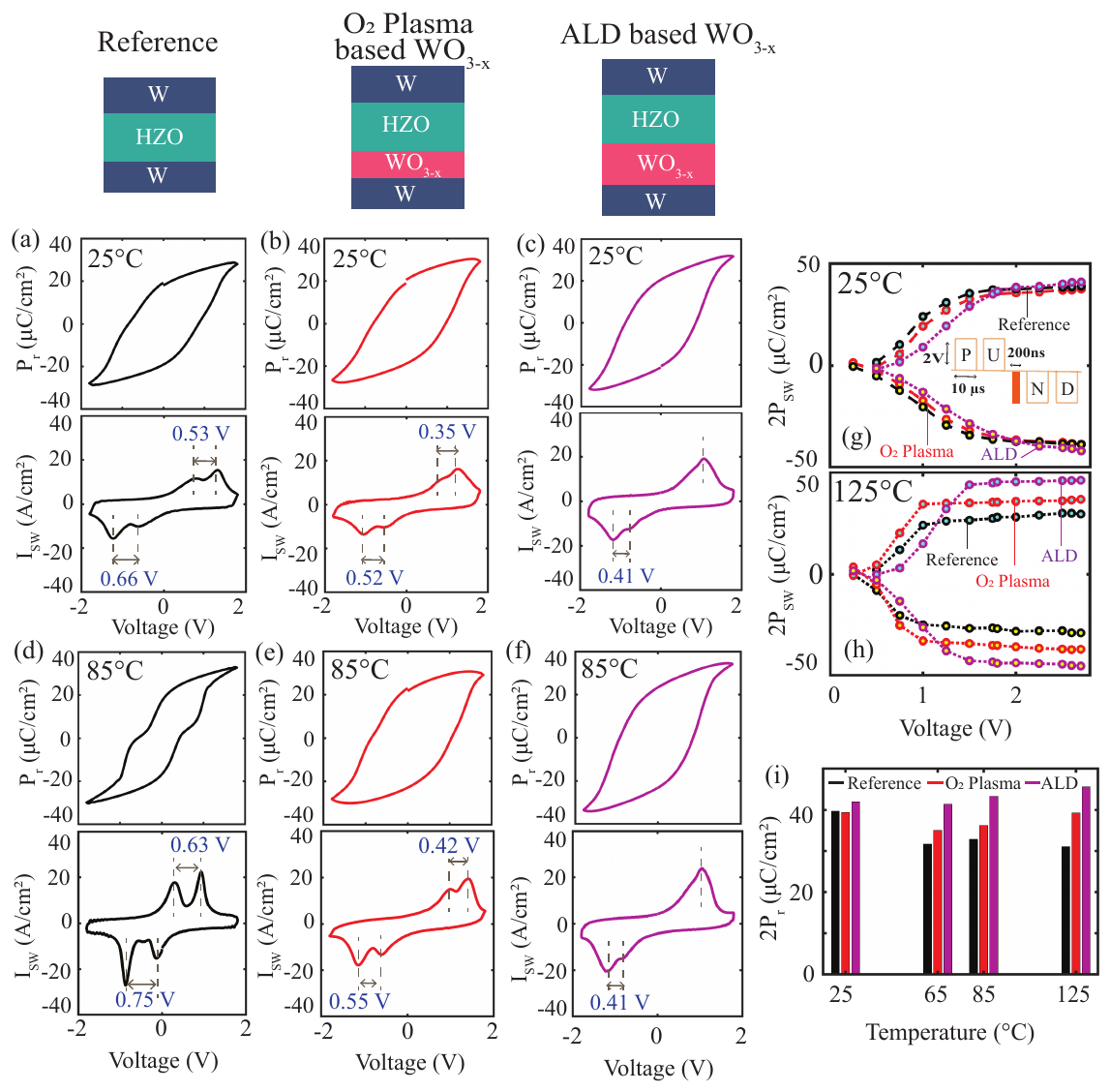}

\caption{Temperature dependent polarization and switching current characteristics at pristine state. P-V and I\textsubscript{SW}-V characteristics at (a-c) 25 and (d-f) 85\textdegree C from (a,d) reference, (b,e) O\textsubscript{2} plasma- and (c,f) ALD- based WO\textsubscript{3-x} devices respectively. (g-h) Polarization switched with 200ns pulses at different voltages measured at 25 and 125\textdegree C respectively on pristine devices. (i) 2P\textsubscript{r} obtained from PUND measurement with 2V/10 $\mu$s square pulses at different temperatures at pristine state. }
\end{figure}

To confirm the stabilization of the ferroelectric o- phase with the incorporation of WO\textsubscript{3-x} at elevated temperatures, grazing incidence X-ray diffraction (GI-XRD) measurements of the HZO layer in reference, O\textsubscript{2} plasma and ALD based WO\textsubscript{3-x} samples were performed at 25, 85, and 125\textdegree C. As shown in Figure 5a, the major o-(111)/t-(101) diffraction peak through Gaussian fitting from the samples having WO\textsubscript{3-x} are shifted toward lower angles compared to the reference device, indicating a greater orthorhombic phase fraction. This observation aligns with the known peak positions of the o-(111) and t-(101) phases in HZO, located at approximately 30.4° and 30.8°, respectively. At 85\textdegree C, the peak separation between these samples decreases; however, the samples having WO\textsubscript{3-x} still exhibit a discernible shift toward the orthorhombic phase, suggesting that a greater proportion of the orthorhombic phase is retained relative to the reference sample (Figure 5b). This behavior persists at 125\textdegree C, as shown in Figure 5c. The phase fractions of the o- and t- phases, extracted from the deconvolution of the o-(111)/t-(101) peaks measured at 125 °C, are shown in Figure S7. The evolution of the peak positions with temperature for all the samples is summarized in Figure 5d.

These results clearly demonstrate that the incorporation of WO\textsubscript{3-x} helps maintain a higher orthorhombic phase fraction across a wide temperature range. And one of the reasons for this could be better lattice matching between orthorhombic Pca2\textsubscript{1} HZO and monoclinic WO\textsubscript{3} compared to that with W (Table S1). Monoclinic phase of WO\textsubscript{3} is chosen for comparison, as this phase is confirmed by STEM imaging of the O\textsubscript{2} plasma device (Figure S8). Additionally, the optimized Vo concentration in ALD-deposited WO\textsubscript{3-x} is likely to promote enhanced o-phase stability in HZO compared to O\textsubscript{2} plasma based WO\textsubscript{3-x}, as prior studies have shown that stabilization of the o- phase in HZO is maximized within an optimal vacancy window in WO\textsubscript{3-x}, whereas both excessively vacancy-rich and near-stoichiometric (vacancy-poor) conditions lead to reduced o-phase content\cite{kim2025x,habibi2025highly}. The enhanced ferroelectric o-phase stability plays a critical role in suppressing the transition to antiferroelectric-like behavior at elevated temperatures, thereby contributing to improved ferroelectric performance.

\begin{figure}[H]
\centering
\includegraphics[ width=6.5in]{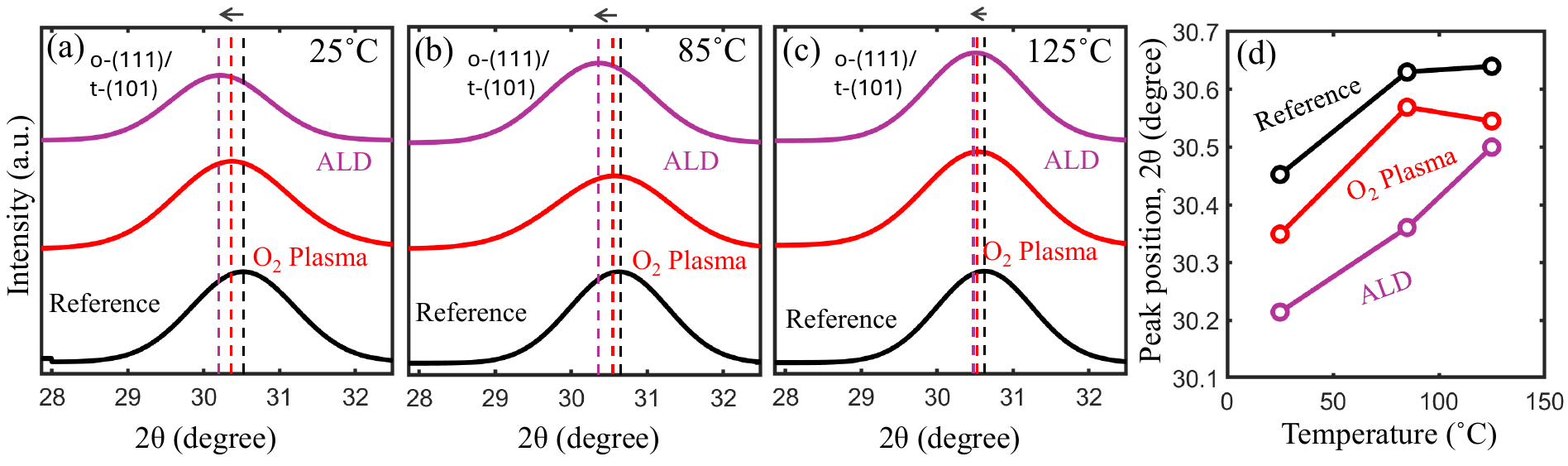}
\caption{Grazing incident X-ray diffraction (GI-XRD) from the HZO film of reference, O\textsubscript{2} plasma and ALD based 6 nm WO\textsubscript{3-x} samples at (a) 25, (b) 85 and (c) 125\textdegree C. (d) Dominant peak (o-111/t-101) position of all the samples at different temperatures obtained from (a)-(c).     }
\end{figure} 

The influence of electrical cycling on the switching behavior, with and without the presence of a WO\textsubscript{3-x} interfacial layer at elevated temperatures, is studied in Figure 6. For the reference device, which lacks WO\textsubscript{3-x}, a significant number of cycles-10\textsuperscript{4} at 85\textdegree C and 10\textsuperscript{5} at 125\textdegree C are required to suppress the characteristic double peaks in the I\textsubscript{SW} response and achieve clean ferroelectric switching (Figures 6a–b). In contrast, the O\textsubscript{2} plasma based WO\textsubscript{3-x} device exhibits stable ferroelectric switching behavior earlier than reference, showing a single peak in the I\textsubscript{SW} profile after just 10\textsuperscript{3} and 10\textsuperscript{4} cycles at 85 and 125\textdegree{C}, respectively (Figures 6c–d).

Remarkably, for the ALD based 6 nm WO\textsubscript{3-x} device, as few as 10 cycles are sufficient for transition to pure ferroelectric switching at both 85 and 125\textdegree C (Figures 6e–f). This phenomena is true for ALD based 5 nm WO\textsubscript{3-x} device as well (Figure S9). The substantial reduction in wake-up cycles observed in ALD-based WO\textsubscript{3-x} compared to O\textsubscript{2} plasma based WO\textsubscript{3-x} devices is consistent with the inherently gentler nature of the ALD process, which proceeds through self-limiting surface reactions and minimizes plasma-induced surface and interfacial defect generation. Collectively, these results underscore the critical role of the WO\textsubscript{3-x} interfacial layer in enhancing the thermal and electrical stability of the ferroelectric phase. The corresponding P–V loops at 85 and 125\textdegree C after different cycling stages are shown in Figures 6(g-l) for reference, O\textsubscript{2} plasma and 6nm ALD WO\textsubscript{3-x} devices and in Figure S10 for 5nm ALD WO\textsubscript{3-x} device, further corroborating these trends. It is worth noting that once pure ferroelectric switching is achieved in the ALD based WO\textsubscript{3-x} devices at high temperatures after 10 bipolar cycles, it remains well preserved upon cooling to room temperature (Figure S11). Table 1 presents a benchmark of the 6 nm ALD WO\textsubscript{3-x} device against prior studies on high-temperature ferroelectric capacitor behavior. The results indicate that the ALD WO\textsubscript{3-x} device achieves superior pristine polarization and undergoes significantly fewer wake-up cycles during elevated-temperature operation compared with recent reports. The 2P\textsubscript{r} values of this work are obtained from the PUND measurement of Figure 4i.

The leakage current density and trap generation rate with electrical cycling of all devices at different temperatures are presented in Figure S12. Although all the devices including reference device show similar trap generation rate at room temperature, WO\textsubscript{3-x} containing devices show a significantly lower trap generation rate during electrical cycling at elevated temperatures, irrespective of the WO\textsubscript{3-x} deposition method (Figure S12(d–f)). This reduced trap generation is attributed to the oxygen-reservoir function of the WO\textsubscript{3–x} layer, which contributes to the improved endurance (as shown in Figure S2) observed in these devices\cite{10983715}.

\begin{figure}[H]
\centering
\includegraphics[ width=6.5in]{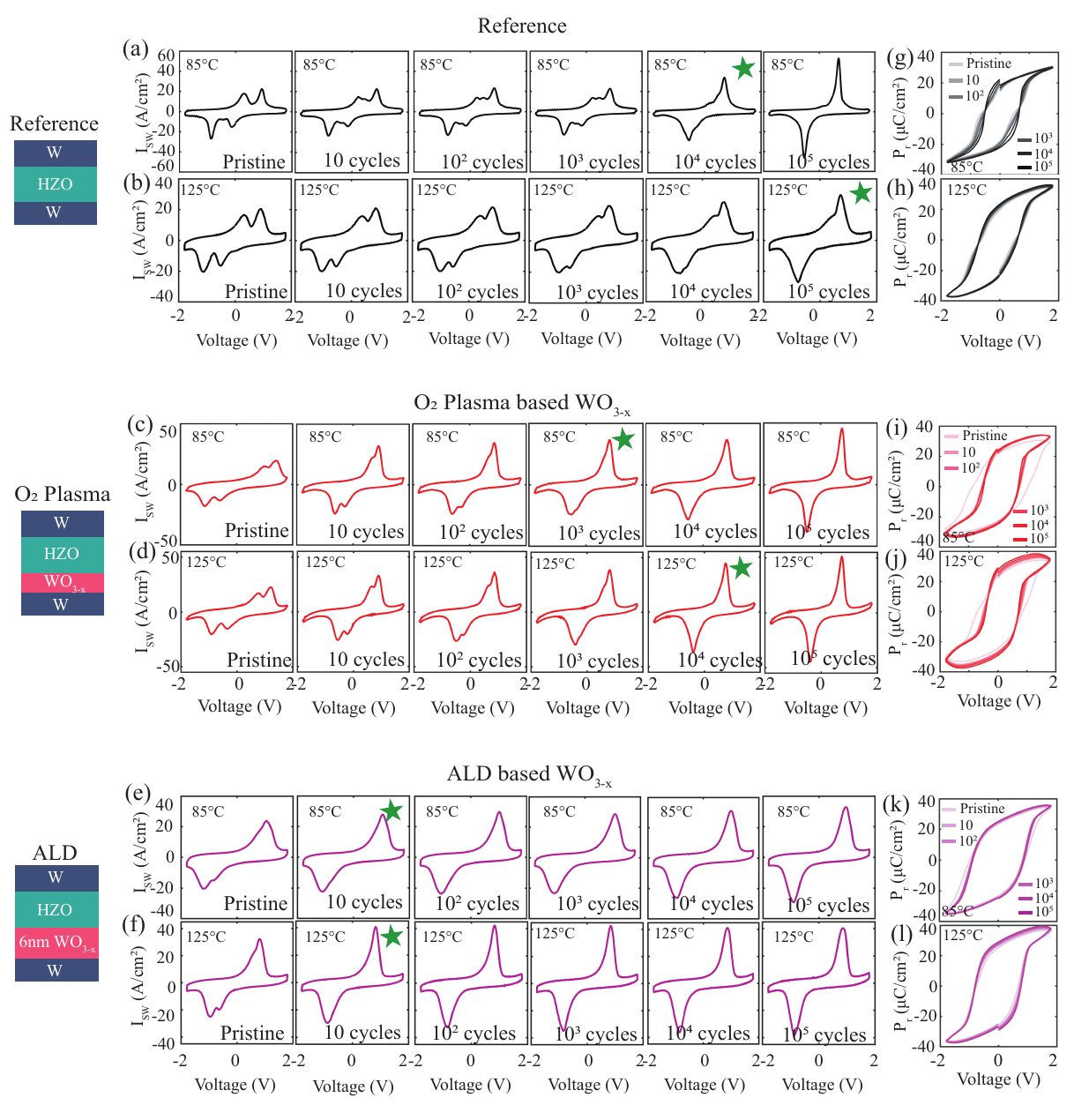}
\caption{Temperature and cycling dependent P-V and I\textsubscript{SW}-V characteristics. I\textsubscript{SW}-V characteristics at (a,c,e) 85 and (b,d,f) 125\textdegree C from pristine to 10\textsuperscript{5} cycles for (a-b) reference, (c-d) O\textsubscript{2} plasma based WO\textsubscript{3-x} and (e-f) ALD based WO\textsubscript{3-x} device. Green asterisks denote the number of cycles which are required to remove the anti-ferroelectric like double peak behavior. P-V loops from pristine to to 10\textsuperscript{5} cycles at (g,i,k) 85 and (h,j,l) 125\textdegree C of (g-h) reference, (i-j) O\textsubscript{2} plasma based WO\textsubscript{3-x} and (k-l) ALD based WO\textsubscript{3-x} device.} 
\end{figure}

\begin{table}[ht]
\centering
\renewcommand{\arraystretch}{1.3}
\begin{adjustbox}{width=\textwidth}
\begin{tabular}{lcccccc}
\toprule
 & \textbf{[29]} & \textbf{[16]} & \textbf{[17]} & \textbf{[25]} & \textbf{[45]} & \textbf{This work} \\
\midrule
Film thickness (nm) & 10 & 10 & 7 & 3 & 12 & \textbf{5} \\
Write Voltage (V) & 3.5 & 2.2 & 2.1 & 1 & 3 & \textbf{1.8} \\
Electric field (MV/cm) & 3.5 & 2.2 & 3 & 3.3 & 2.5 & \textbf{3.6} \\
Pristine 2$P_r$ ($\mu$C/cm$^2$) & 26 & 15.1 & 16 & 5 & 32.94 & \textbf{41.94} \\
Pristine 2$P_r$ at HT ($\mu$C/cm$^2$) & -- & 7 (120$^\circ$C) & 13.56 (125$^\circ$C) & 11.81 (85$^\circ$C) & 24.95 (100$^\circ$C) & \textbf{45.57 (125$^\circ$C)} \\
Cycles to wakeup at HT & 10$^3$ (100$^\circ$C) & 10$^6$ (100$^\circ$C) & 10$^4$ (100$^\circ$C) & 10$^3$ (85$^\circ$C) & 10$^6$ (100$^\circ$C) & \textbf{10 (125$^\circ$C)} \\
\bottomrule
\end{tabular}
\end{adjustbox}
\caption{Comparison of ferroelectric properties at high temperature (HT) with prior works.}
\end{table}

\subsection{Density functional theory (DFT)}
To investigate the origin of the stabilization of ferroelectric properties at elevated temperatures in the presence of WO\textsubscript{3-x}, density functional theory (DFT) calculations were performed to compare the stability of the orthorhombic Pca2\textsubscript{1} phase in HZO when deposited on W and WO\textsubscript{3}. In this analysis, the atomic structure of bulk HZO was strained to match the lattice parameters of W and WO\textsubscript{3}, as illustrated in Table S1. Although the Tungsten oxide layer in the fabricated ferroelectric capacitors is sub-stoichiometric, HZO lattice is strained to WO\textsubscript{3} lattice for the calculation as the dominant feature in the W 4f XPS spectra corresponds to W\textsuperscript{6+} oxidation state. Furthermore, WO\textsubscript{3-x} containing oxygen vacancies does not exhibit a single well-defined periodic lattice constant due to the configurational disorder introduced by these vacancies. To account for oxygen deficiency, lattice mismatch calculations were carried out using WO\textsubscript{3} supercell containing a single oxygen vacancy, as summarized in Table S1. The mismatch values indicate that oxygen vacancies can tune the cell-averaged mismatch while primarily introducing local strain inhomogeneity\cite{wang2016role}. However, the key thermodynamic and kinetic trends discussed henceforth remain qualitatively unchanged.

Figure 7a presents the Helmholtz free energy of HZO in the monoclinic (m-), tetragonal (t-), and o- phases, each strained to match the lattice parameters of W and WO\textsubscript{3}, respectively. The yellow-shaded region indicates the high operating temperature range of interest (65–125 \textdegree C). Within this range, the energy difference between the t- and o-phases of HZO strained to the WO\textsubscript{3} lattice increases by approximately 20\% compared to that strained to the W lattice, thereby favoring formation of the o-phase in the WO\textsubscript{3} case. This behavior arises because W, having a cubic structure, induces a more cubic-like configuration in HZO, resulting in higher entropy relative to the WO\textsubscript{3} lattice (as reflected by the slope of the Helmholtz free energy curve). Consequently, the energy gap between the t- and o-phases is larger for WO\textsubscript{3} at elevated temperatures. While W exhibits a greater lattice mismatch and thus higher surface energy than WO\textsubscript{3}, the increase in surface energies for both the o- and t-phases is comparable, yielding only a minor overall effect on the Helmholtz free energy.

Although the m-phase remains the thermodynamically most stable state even at a film thickness of 5 nm (i.e., exhibiting the lowest Helmholtz free energy), the kinetic barrier for the t → o phase transition is substantially lower than that for the t → m transition. Consequently, under kinetic constraints, the system preferentially transforms into the o-phase rather than the m-phase \cite{10.1063/5.0160719}, in agreement with the calculations presented in Table S2. During cooling, the high energy barrier similarly suppresses the formation of the m-phase, leading to the preferential development of the o-phase, which possesses a lower Helmholtz free energy than the t-phase. This trend is further supported by Figure 7a, where the o-phase remains more stable than the t-phase even at elevated operating temperatures, making it the energetically favorable configuration.

As shown in Figure 7b, t-phase of HZO exhibits only the $X\textsubscript{2}'$ oxygen phonon mode, which drives the cubic-to-tetragonal phase transition. When strain is applied to match the lattice constant of WO\textsubscript{3}, corresponding to an approximately 1.7\% increase in the lattice constant along the x-direction relative to the pristine structure, the $X\textsubscript{2}'$ phonon mode in the t-phase is enhanced. This enhancement brings its value closer to that of the o-phase, indicating that the t → o phase transition becomes more accessible. Under W lattice, the larger lattice mismatch produces a greater amplitude of the $X\textsubscript{2}'$ mode compared to the WO\textsubscript{3} case, leading, as shown in Table S2, to a lower t → o transition energy barrier. Overall, while the larger strain induced by W reduces the transition barrier relative to WO\textsubscript{3}, the o-phase in the WO\textsubscript{3} case remains thermodynamically more stable than the t-phase due to the combined effects of surface energy at the 5 nm thick HZO and the large entropy associated with the larger in-plane lattice matched to the W cubic structure under non-equibiaxial strain (as described in Figure S13). These results are consistent with previous phase stabilization studies for the ZrO\textsubscript{2}–WO\textsubscript{3} interface\cite{zhao2023engineering}.

The polarization switching barrier of HZO strained to the W and WO\textsubscript{3} lattices were further calculated and found to be lower in the WO\textsubscript{3}-strained case (Figure S14). This difference arises because, in the switching pathway, the transition state corresponds to the t-phase, which exhibits a longer lattice constant along the x-direction than along the y-direction. For W, the structure is more cubic-like, with nearly identical lattice constants in both directions, leading to a higher switching barrier. In contrast, the anisotropy in the WO\textsubscript{3}-strained lattice reduces the barrier height. This energy barrier difference contributes to improved overall switching kinetics in devices incorporating WO\textsubscript{3} or WO\textsubscript{3-x} as the interfacial layer.

\begin{figure}[H]
\centering
\includegraphics[ width=6.5in]{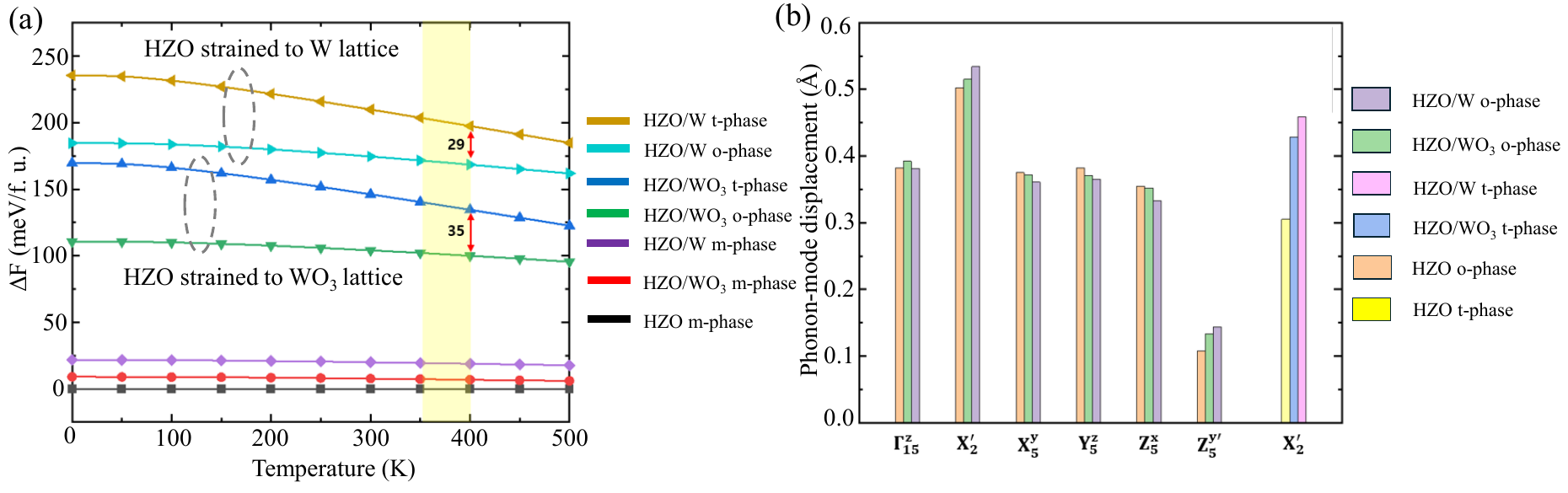}
\caption{Density functional theory (DFT) analysis. (a) Helmholtz free energy difference ($\Delta$F) for the 5 nm HZO film, relative to the m- phase. The yellow region and red arrow indicate the high-temperature operating range and free energy difference between t- and o- phase under W and WO\textsubscript{3}-induced strain. (b) Phonon mode analysis under tensile strain induced by W and WO\textsubscript{3}, where the enhancement of the soft $X\textsubscript{2}'$ mode in the t-phase induces the phase transition to the o-phase.
}
\end{figure}

\section{Conclusion}
In this study, we demonstrated the critical role of WO\textsubscript{3-x} interfacial layer in enhancing the thermal stability and ferroelectric performance of ultrathin Hf\textsubscript{0.5}Zr\textsubscript{0.5}O\textsubscript{2} (HZO)-based capacitors. The incorporation of WO\textsubscript{3-x}, regardless of the deposition method and with carefully tuned oxygen vacancy, effectively suppressed the transition toward antiferroelectric-like behavior at elevated temperatures and enabled nearly wakeup-free operation with significantly fewer bipolar cycles. This improvement is attributed to the reduced relative Helmholtz energy for the orthorhombic phase in the presence of WO\textsubscript{3-x}, which favors the retention of ferroelectric properties even under thermal stress. These findings offer a promising pathway toward the realization of robust ferroelectric memories suitable for 3D memory-on-logic architectures, where reliable high-temperature operation is essential.

\section{Methods}
\subsection{Device fabrication}
Ferroelectric HZO capacitors were fabricated on p\textsuperscript{+} Si(100) substrates with a doping concentration of 10\textsuperscript{20} cm\textsuperscript{-3}. A 20 nm W bottom electrode was deposited using a Unifilm sputter system at a deposition rate of 200 \text{\AA}. 5 nm HZO layer was deposited using the Kurt J. Lesker ALD tool at 250\textdegree C using TDMAH and TDMAZ precursors for Hf and Zr, and H\textsubscript{2}O as oxidant in the Th-ALD process -directly on W in case of reference sample. For the O\textsubscript{2} plasma sample, the W electrode was oxidized to form a WO\textsubscript{3-x} interfacial layer through 10 cycles of Oxygen plasma treatment in the ALD chamber prior to HZO deposition, ensuring no vacuum break. For the ALD WO\textsubscript{3-x} samples, 20 and 30 cycles of ALD were used to deposit 5 nm and 6 nm WO\textsubscript{3-x} layers respectively before depositing HZO in Veeco Fiji G1 ALD system with a Bis(tert-butylimino)bis(dimethylamino) (BTBMW) tungsten precursor and O\textsubscript{2} plasma (300 W power, 40 sccm O\textsubscript{2} flow, and 5s plasma exposure per cycle) as oxidant at 250\,\textdegree C. 70 nm W top electrode was sputtered onto all devices, followed by rapid thermal annealing (RTA) at 550\textdegree C for 1 minute in N\textsubscript{2} atmosphere. Finally, standard photolithography and dry etching were used to define $50~\mu\text{m} \times 50~\mu\text{m}$ capacitor structures.

\subsection{Energy-dispersive X-ray spectroscopy (EDS)}
 The EDS maps were collected using a Bruker E3DS detector with a 60 mm\textsuperscript{2} window, on cross-sectional focused ion beam (FIB)-prepared samples in a Hitachi HD-2700 STEM. The spectra was then read via Hyperspy and background subtractions were performed by selecting one window before and one window after the peaks of interest. Peaks corresponding to Zr-K$\alpha$, W-L$\alpha$, and O-K$\alpha$ were used for mapping. To denoise the spectra, principal component analysis (PCA) was first applied, retaining four components, and then Online Robust Nonnegative Matrix Factorization (ORNMF) was performed using these four components identified. Line scan were created at the center of horizontal direction (x axis) for each map with a width of 40 pixels.

\subsection{X-ray photoelectron spectroscopy (XPS)}
X-ray photoelectron spectroscopy (XPS) was conducted after depositing WO\textsubscript{3-x} (either by ALD or using O\textsubscript{2} plasma) on sputtered bottom W electrode on Si substrate. Thermo K-alpha XPS System with an Al K$\alpha$ x-ray source and a beam spot size of 200 µm was used for the scan. Flood gun was always on during the data acquisition. XP emission peaks are charge corrected to the carbon 1s peak which is set at 284.8 eV. XP emission backgrounds are subtracted with Shirley algorithm, and the peaks are fitted with Powell algorithm, with convergence level < 0.0001.

\subsection{Grazing incident X-ray diffraction (GI-XRD)}
After removing top electrode of the devices using W etchant, GIXRD scan was done using a RIGAKU Smartlab XE diffractometer equipped with a Cu K\(\alpha\) source (40kV, 50mA) and a HyPix-3000HE detector. In-situ high temperature data is collected on RIGAKU Reactor X stage. Data was acquired in the range of 27°-34°, with an incidence angle of 0.5°, scanning step of 0.04°, and scanning speed of 0.05°/min. O and T peaks are constrained to have gaussian shape and same FWHM when deconvolution.

\subsection{Electron Microscopy}
Cross-sectional samples for scanning transmission electron microscopy (STEM) imaging and EDS were prepared using a Thermo Fisher Helios 5CX FIB/SEM equipped with a high energy Focused Ion Beam (FIB) using Gallium-69 and operated at an accelerating voltage between 0.1 and 30 kV. The final polishing condition was performed at 10 pA at 2kV. A Hitachi HD-2700 aberration- corrected STEM/SEM was used to capture STEM images, operated at a 200 kV accelerating voltage and 27 mrad convergence semi-angle. The spatial resolution was about 1.3 \AA.

\subsection{Electrical measurements}
Electrical measurements were performed using a Cascade Microtech Summit 1200K semi-automated probe station, equipped with a Keysight B1500 semiconductor device analyzer. A 20 $\mu\text{s}$ triangular pulse with an amplitude of 1.8 V was applied to measure the P–V and I$_{SW}$-V characteristics. For bipolar cycling, 200 ns pulses with an amplitude of 1.8 V were used. PUND measurements were conducted using 10 $\mu\text{s}$ trapezoidal pulses, each with 10 $\mu\text{s}$ rise and fall times. All measurements were done on $50~\mu\text{m} \times 50~\mu\text{m}$ devices.

\subsection{Density functional theory (DFT)}
Density functional theory (DFT) calculations were performed using the Vienna Ab initio Simulation Package (VASP)\cite{kresse1996efficient, kresse1993ab, kresse1996efficiency, kresse1999ultrasoft}. The electron–core interactions were described using the local density approximation (LDA)\cite{ceperley1980ground, perdew1981self} in conjunction with Blöchl’s projector augmented wave (PAW) method\cite{kresse1999ultrasoft,blochl1994projector}. A plane-wave cutoff energy of 500 eV was employed, and k-point meshes were sampled using the Monkhorst-Pack (MP)\cite{monkhorst1976special} method with an 8×8×8 for conventional unit cells.  Atomic positions were fully relaxed until the total energy and interatomic forces converged to less than 10–8 eV and 0.01 eV/\AA, respectively. Vibrational free energies were obtained using the finite displacement method as implemented in Phonopy\cite{togo2015first}, which enabled accurate computation of Helmholtz free energy. The surface energies were calculated using slab models comprising more than 8 layers with a vacuum region thicker than 15\AA. For slab calculations, the k-point meshes of MP method 6×6×1 were used. The film thickness was set to 5 nm.
The Helmholtz free energy F was calculated using the following relation :
\[
F = U + F\textsuperscript{vib} + \gamma \Omega
\]
where U is the internal energy of bulk system, F\textsuperscript{vib} is the vibrational free energy, $\gamma$ is the surface energy, and $\Omega$ is the surface area.

\section*{Supporting Information}
XPS of 5nm ALD based WO\textsubscript{3-x}; Endurance; Imprint; Retention; Conductivity; Pristine P-V and I\textsubscript{SW}-V characteristics of 5nm ALD based WO\textsubscript{3-x} device; Deconvoluted GI-XRD peaks of 125\textdegree C measurement; Lattice mismatch\% of HZO with W, WO\textsubscript{3} and single vacancy-contained WO\textsubscript{3}; STEM image of WO\textsubscript{3}; I\textsubscript{SW}-V and P-V of 5nm ALD WO\textsubscript{3-x} device at 85 and 125\textdegree C; I\textsubscript{SW}-V of 6nm ALD based WO\textsubscript{3-x} device when cooled back to RT from 85 and 125\textdegree C; Leakage and trap generation rate with cycling; Energy barrier of HZO phase transition; Vibrational entropies of W and WO\textsubscript{3}; Switching barrier energy with and without WO\textsubscript{3}.

\section*{Author Contribution}

N.A and J.C. contributed equally to this work. N.A. designed and performed the experiments and analyzed the results. J.C. and C.H.K. performed DFT calculations and analysis. S.S and Y.-H.K fabricated the devices. S.S., J.C., M.T., C.Z. and P.G.R. assisted with the experiments and data analysis. A.K. and J.H.L. supervised the research. N.A. and J.C. wrote the manuscript with the inputs from all the authors.

\begin{acknowledgement}
This work involving reference and O\textsubscript{2} plasma WO\textsubscript{3-x} devices was supported by SUPREME, one of the seven SRC-DARPA JUMP 2.0 centers. The work with ALD WO\textsubscript{3-x} devices was supported by the Center for 3D Ferroelectric Microelectronics Manufacturing, an Energy Frontier Research Center funded by the U.S. Department of Energy, Office of Science, Basic Energy Sciences, under Award No. DE-SC0021118. Fab was done at the IMS at Georgia Tech, supported
by the NSF-NNCI program (ECCS-1542174). DFT work was supported by National Research Foundation of Korea (RS-2023-00218799, RS-2024-00404361, RS-2023-00257666, RS-2025-24535610), Industrial Technology Innovation Program (RS-2025-06642983), Korea Institute for Advancement of Technology (KIAT) grant funded by the Korea Government (MOTIE) (P0023703, HRD Program for Industrial Innovation) and National Supercomputing Center with supercomputing resources including technical support (KSC-2022-CRE-0075, KSC-2022-CRE-0454, KSC-2022-CRE-0456, KSC-2023-CRE-0547, KSC-2024-CRE-0545).

\end{acknowledgement}

\providecommand{\latin}[1]{#1}
\makeatletter
\providecommand{\doi}
  {\begingroup\let\do\@makeother\dospecials
  \catcode`\{=1 \catcode`\}=2 \doi@aux}
\providecommand{\doi@aux}[1]{\endgroup\texttt{#1}}
\makeatother
\providecommand*\mcitethebibliography{\thebibliography}
\csname @ifundefined\endcsname{endmcitethebibliography}  {\let\endmcitethebibliography\endthebibliography}{}

\newpage
\makeatletter
\setcounter{figure}{0}
\renewcommand{\thefigure}{S\arabic{figure}}
\renewcommand{\figurename}{Figure}
\renewcommand{\fnum@figure}{\figurename~\thefigure}
\makeatother

\setcounter{table}{0}
\renewcommand{\thetable}{S\arabic{table}}

\title{\vspace{-1.0em}\Large Supporting Information\\[0.4em]

Figure S1 shows the XPS spectra coming from 5nm ALD deposited WO\textsubscript{3-x} layer on W bottom electrode and Si substrate. W 4f spectrum shows dominant W\textsuperscript{6+} peaks confirmning the formation of WO\textsubscript{3-x}. Deconvoluted peaks of O 1s spectrum are shown in Figure S1b.
 
\begin{figure}[H]
\centering
\includegraphics[ width=6.5in]{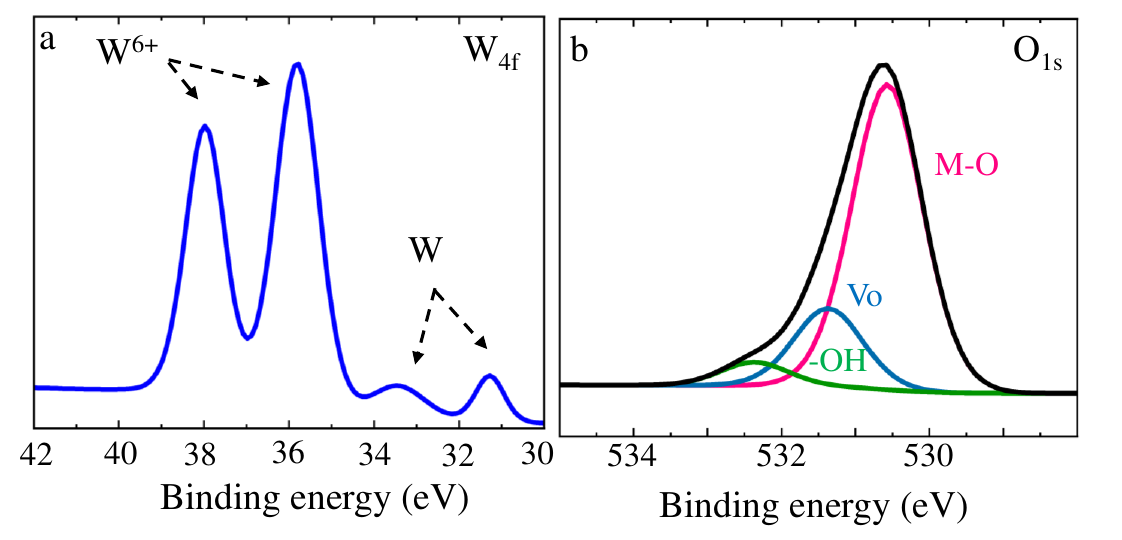}
\caption{XPS spectra obtained from (a) W 4f and (b) O 1s orbitals of WO\textsubscript{3-x} from 5 nm ALD grown WO\textsubscript{3-x} samples. (b) Magenta, blue and green curves are the de-convoluted peaks corresponding to M-O, non-lattice Oxygen (V\textsubscript{o}) and -OH respectively.   }
\label{XPS_5nm}
\end{figure}

Figure S2a shows 2P\textsubscript{r} versus cycles characteristics at room temperature when 200 ns bipolar pulses of 1.8 V were applied. The reference device lacking WO\textsubscript{3-x} has poor endurance compared to both O\textsubscript{2} plasma and ALD based WO\textsubscript{3-x} devices. WO\textsubscript{3-x} devices didn't break upto 10\textsuperscript{10} and 10\textsuperscript{11} cycles. The high-temperature endurance characteristics are presented in Figure S2b. Both the O\textsubscript{2} plasma and ALD WO\textsubscript{3–x} devices exhibit higher endurance than the reference device over a wide temperature range. Although the ALD WO\textsubscript{3–x} device reaches saturation polarization at 1.8 V (Figure 4g), endurance measurements were also performed using ±2 V bipolar pulses (orange curve) because of its slightly higher coercive voltage. At least three devices were measured for each condition.

\begin{figure}[H]
\centering
\includegraphics[ width=6.5in]{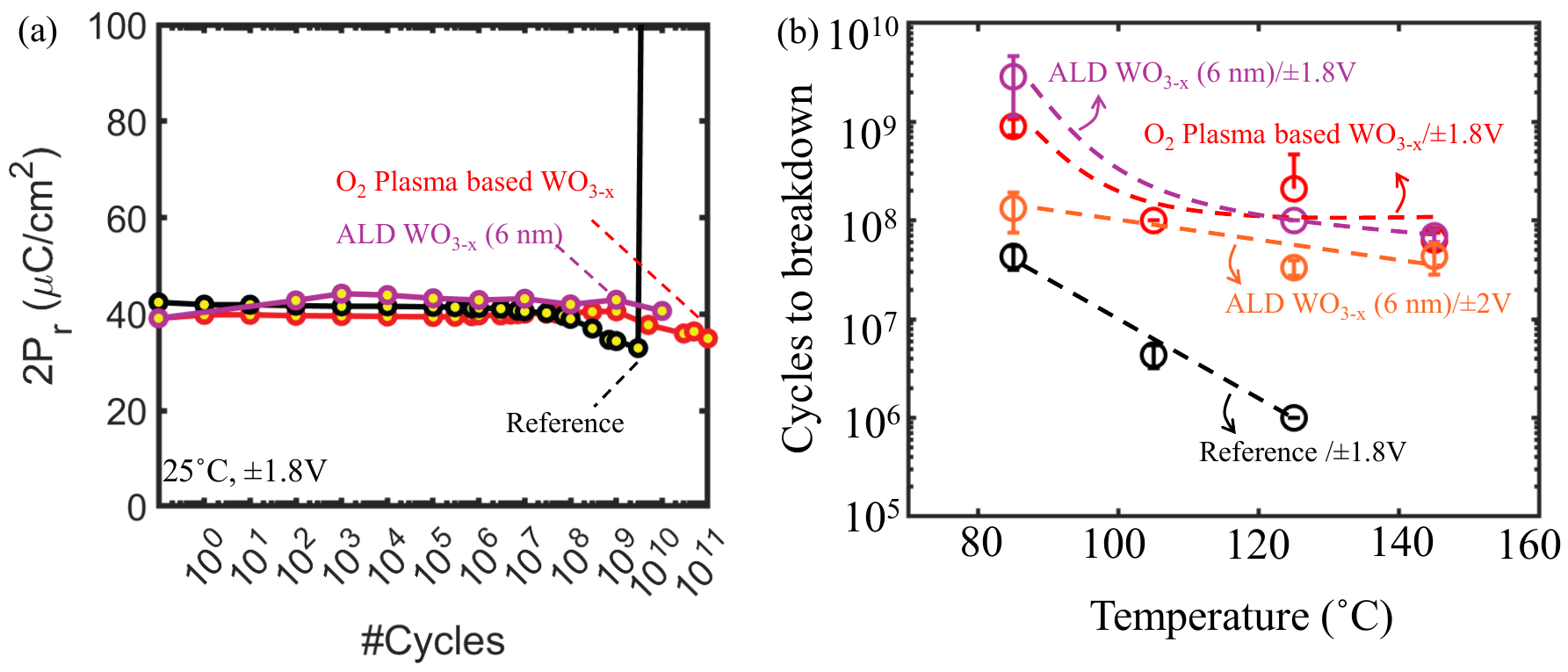}
\caption{(a) 2P\textsubscript{r} versus cycles characteristics with bipolar fatigue pulses of ±1.8 V/200 ns. PV measurements were also done at ±1.8 V. (b) Cycles to breakdown (endurance) at high temperatures. Both ±1.8 V and ±2 V bipolar cycling results are shown for ALD 6nm WO\textsubscript{3-x} devices.}   
\label{Endurance}
\end{figure} 

Figure S3 shows coercive voltage (V\textsubscript{c}) versus cycles characteristics at 125\textdegree C when 200ns bipolar pulses of 1.8V were applied. Imprint doesn't degrade due to the presence of WO\textsubscript{3-x} compared to reference device.

\begin{figure}[H]
\centering
\includegraphics[ width=4.5in]{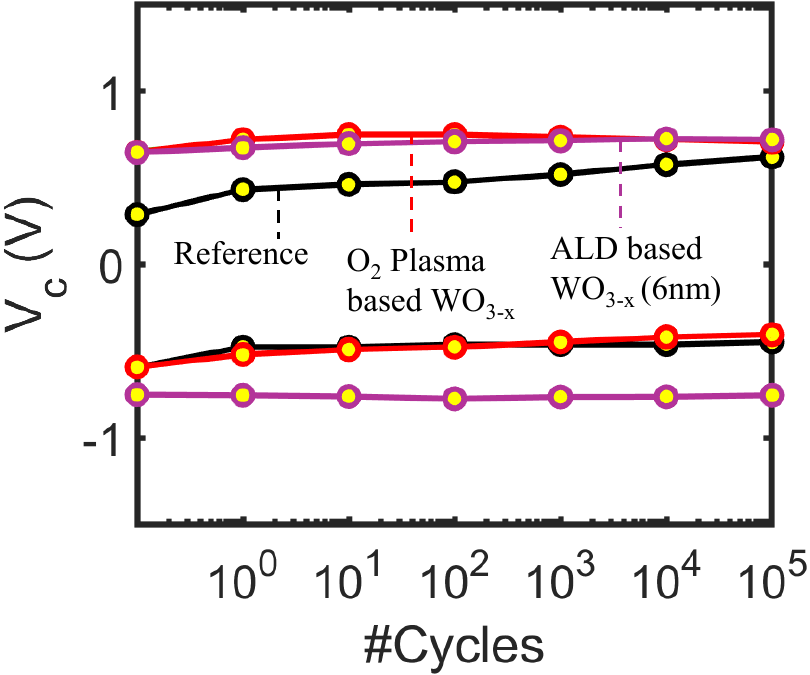}
\caption{Shift in coercive voltage versus cycles at ±1.8V/200 ns , 125\textdegree C.}   
\label{Imprint}
\end{figure} 

Figure S4 shows retention characteristics at 85\textdegree C by applying ±1.8 V square pulses. Retention doesn't degrade due to the presence of WO\textsubscript{3-x} compared to reference device.

\begin{figure}[H]
\centering
\includegraphics[ width=4.5in]{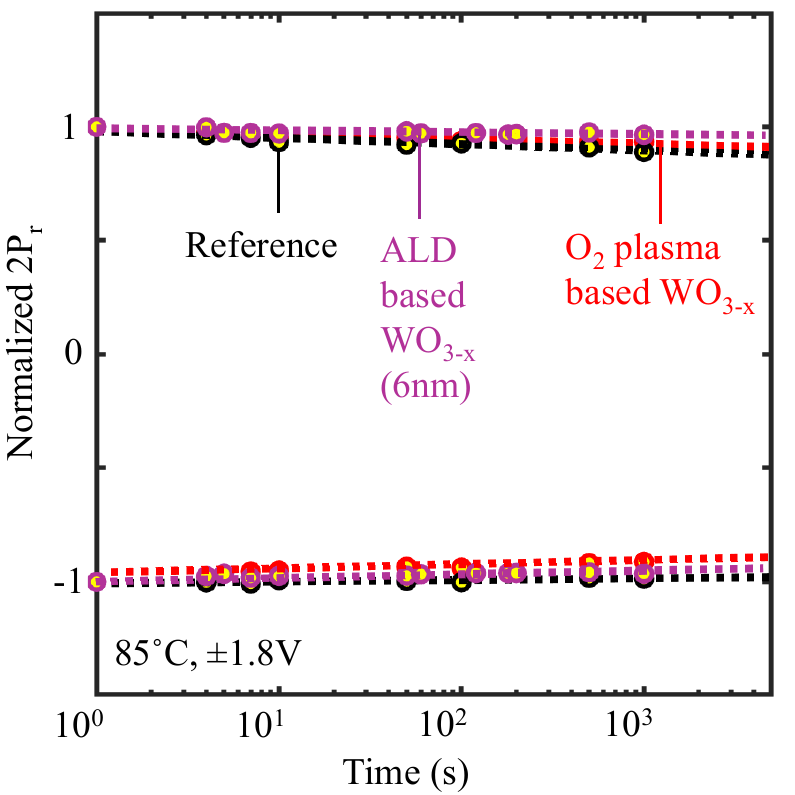}
\caption{Retention characteristics at 85\textdegree C with ± 1.8 V.}   
\label{Imprint}
\end{figure}

Figure S5 shows conductivity of O\textsubscript{2} plasma and ALD based 6 nm WO\textsubscript{3-x} devices measured on 50$\mu$mx50$\mu$m devices by applying AC voltage of 100kHZ, 25 mV and DC sweep of 1.2 V.

\begin{figure}[H]
\centering
\includegraphics[ width=4.5in]{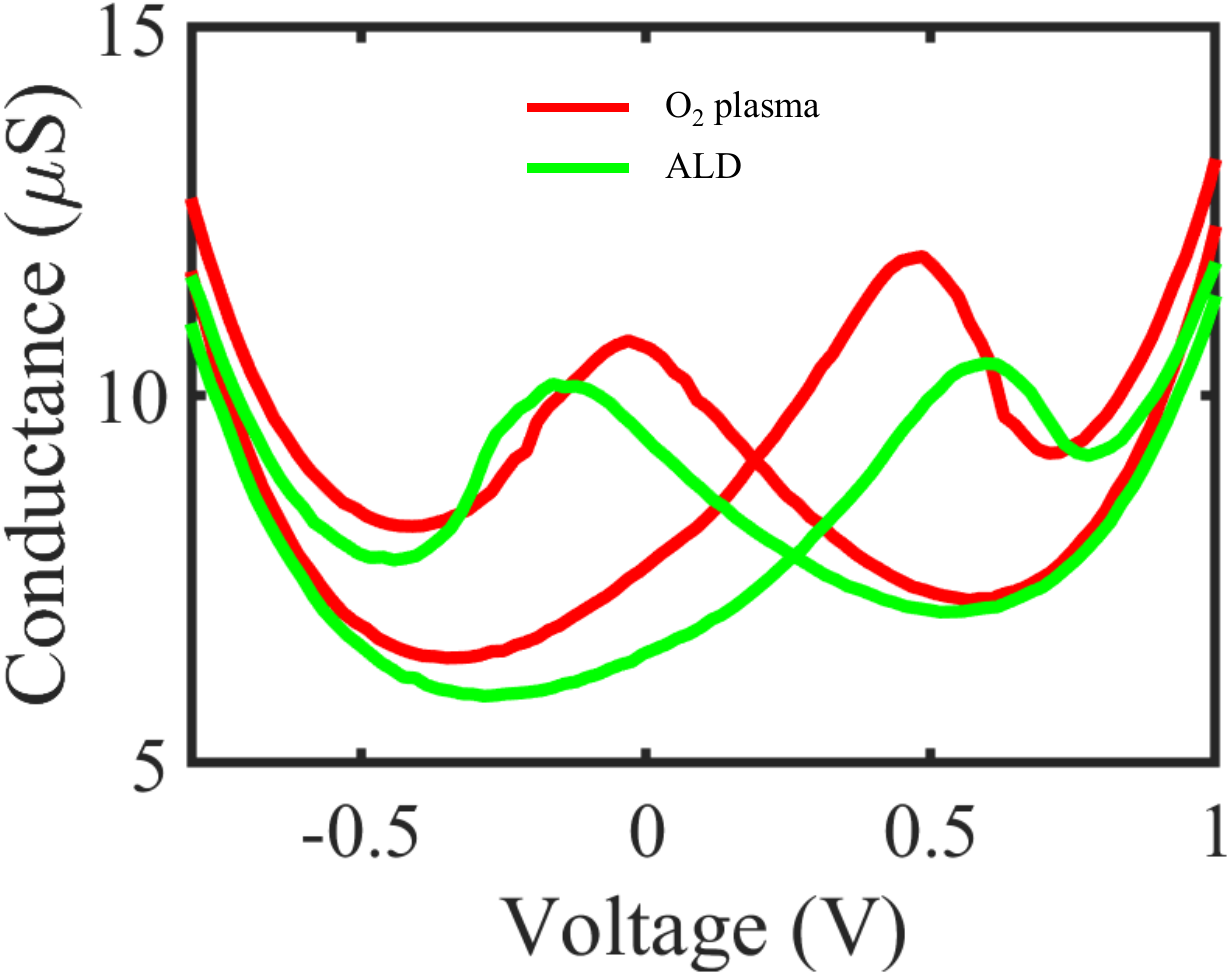}
\caption{Conductivity versus applied voltage. O\textsubscript{2} plasma based WO\textsubscript{3-x} device is more conductive compared to ALD based WO\textsubscript{3-x} device.   }
\label{Conductivity}
\end{figure} 

Figure S6 shows the P-V and I\textsubscript{SW}-V characteristics of pristine device measured at different temperatures from 5nm ALD deposited WO\textsubscript{3-x} device. It shows minimal double peak characteristic (anti-ferro nature) like the 6nm WO\textsubscript{3-x} device. These measurements were done by applying 1.8V/$20\mu$s bipolar triangular pulses.

\begin{figure}[H]
\centering
\includegraphics[ width=6.5in]{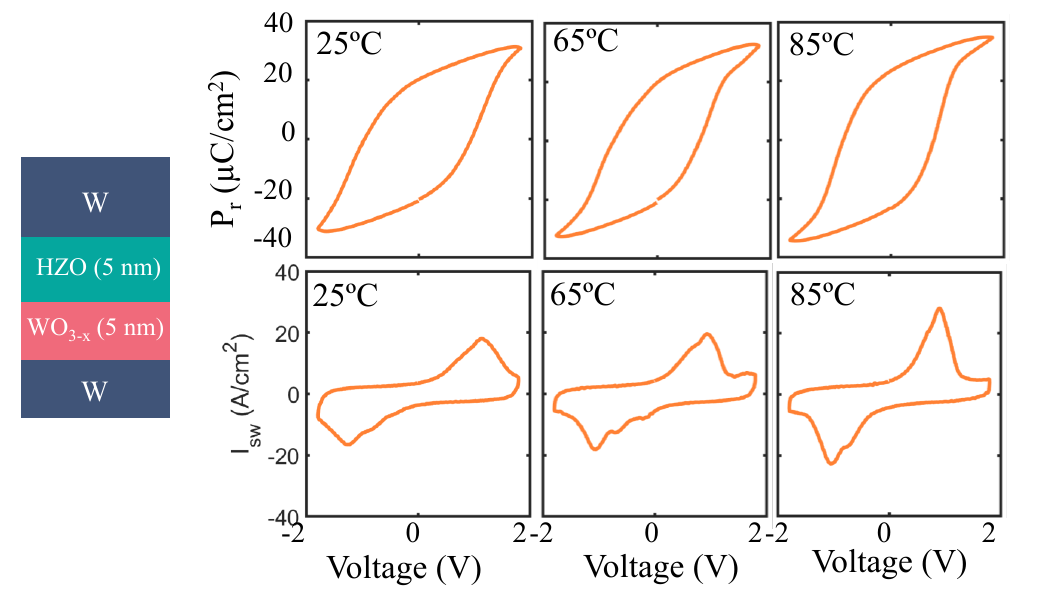}
\caption{ P-V and I\textsubscript{SW}-V characteristics at 25, 65 and 85\textdegree C from pristine ALD deposited 5nm WO\textsubscript{3-x} device.   }
\label{RT_5nm}
\end{figure}

Figure S7 presents the deconvolution of the orthorhombic and tetragonal phase components from the Gaussian-fitted o-(111)/t-(101) diffraction peak measured from the HZO layer at 125\textdegree C. The ALD-deposited WO\textsubscript{3-x} sample exhibits the highest orthorhombic phase fraction (73\%), whereas the reference sample shows the lowest orthorhombic phase content.

\begin{figure}[H]
\centering
\includegraphics[ width=6.5in]{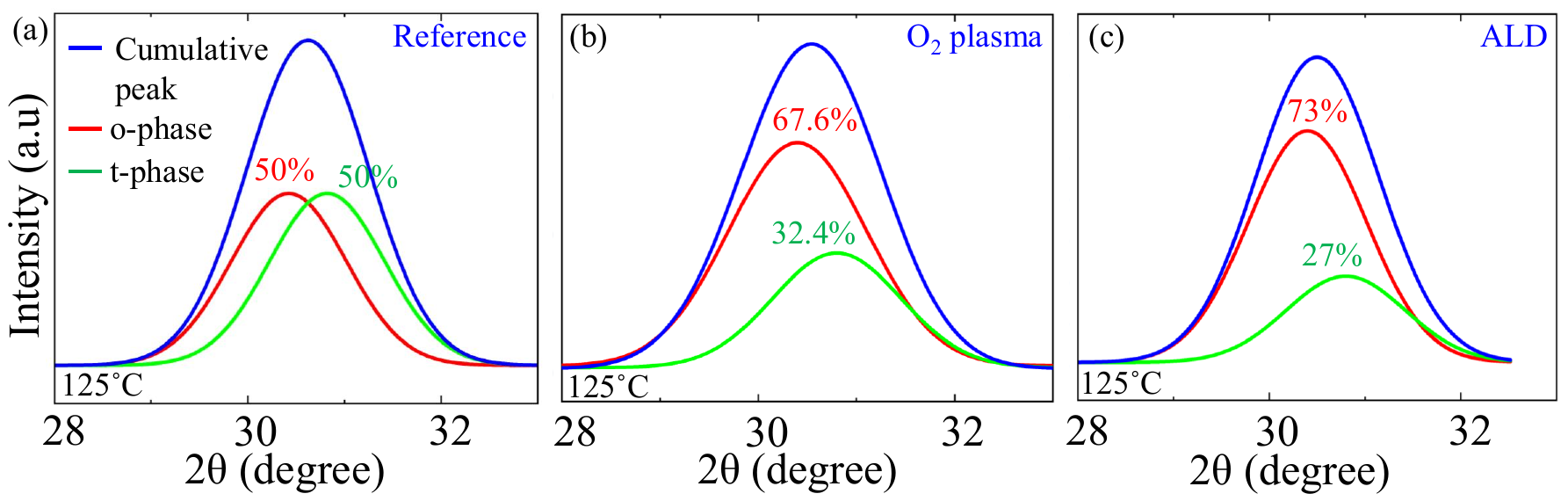}
\caption{Deconvoluted o- (red) and t- (green) phase peaks from the fitted o-(111)/t-(101) diffraction peak (blue) of the HZO layer for (a) reference, (b) O\textsubscript{2} plasma, and (c) 6nm ALD-deposited WO\textsubscript{3-x} samples measured at 125 °C.}
\label{RT_5nm}
\end{figure}

Table S1 shows lattice constant mismatch of HZO Pca2\textsubscript{1} (001) and (111) planes with W, WO\textsubscript{3} and 2x2x2 WO$_3$ supercell containing single Vo\textsuperscript{2+} respectively. The mismatch is lower with both WO\textsubscript{3} and WO\textsubscript{3-x} compared to W in both a and b directions. Supercell containing vacancy can tune the cell-averaged mismatch while primarily introducing local strain inhomogeneity\cite{wang2016role}. Since HZO (111) plane exhibits a hexagonal geometry, the lattice was matched along the x-axis, and an equivalent magnitude of strain, corresponding to that applied along the x-axis, was imposed along the y-axis to introduce in-plane strain.

\begin{table}[h!]
\centering
\renewcommand{\arraystretch}{1.4}
\begin{tabular}{|c|c|c|c|c|c|c|}
\hline
 & W (a) & W (b) & WO$_3$ (a) & WO$_3$ (b) & WO$_3-x$ (a) & WO$_3-x$ (b) \\
\hline
HZO Pca2$_1$ (001) & 1.76\% & 5.84\% & 1.40\% & 4.07\% & 1.21\% & 2.88\% \\
\hline
HZO Pca2$_1$ (111) & 2.56\% &  & 2.48\% & & 2.24\% & \\
\hline
\end{tabular}
\caption{Lattice mismatch (\%) of HZO Pca2$_1$ orientations with W, WO$_3$ and WO$_3$ supercell containing single Vo\textsuperscript{2+}.}
\label{tab:lattice-mismatch}
\end{table}

Figure S8 shows the HAADF image of WO\textsubscript{3} layer of O\textsubscript{2} plasma device. Fast fourier transform (FFT) of the on-zone grain of WO\textsubscript{3} confirms it to be m-(001). Overlayed golden W atoms are coming from the crystal structure of monoclinic WO\textsubscript{3} obtained from .cif file of materials project.

\begin{figure}[H]
\centering
\includegraphics[ width=5.5in]{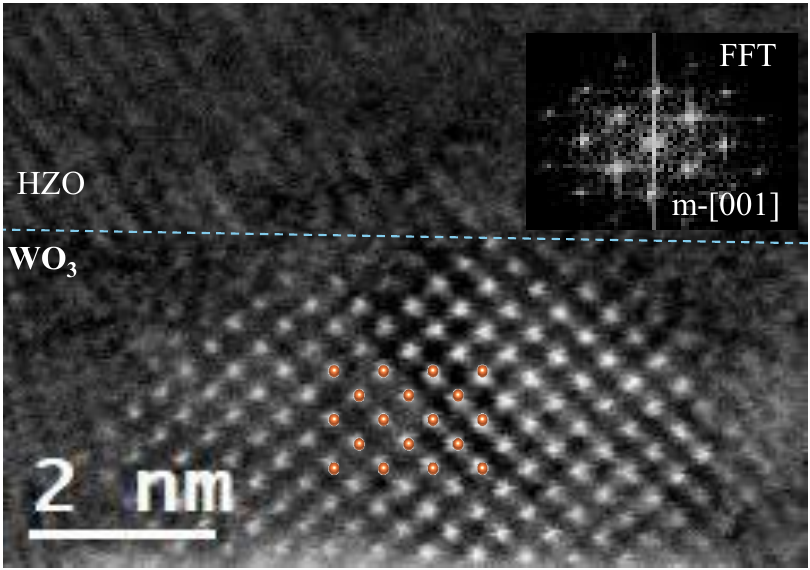}
\caption{STEM image of WO\textsubscript{3} layer of O\textsubscript{2} plasma based device. Golden Overlayed W atoms are obtained from m-(001) crystal model. Fast Fourier Transform (FFT) of the WO\textsubscript{3} grain confirms it to be m-phase with [001] zone axis.}
\label{STEM}
\end{figure} 

Figure S9 shows I\textsubscript{SW}-V characteristics of 5nm ALD deposited WO\textsubscript{3-x} device measured at 85 and 125\textdegree C. Bipolar cycling was done by applying 200ns/1.8V pulses. Double peak in this device goes away just by applying 10 cycles, similar to 6nm ALD based WO\textsubscript{3-x} device. Much less cycles are needed for wakeup in devices having WO\textsubscript{3-x} compared to reference device irrespective of WO\textsubscript{3-x} thickness. 

\begin{figure}[H]
\centering
\includegraphics[ width=6.5in]{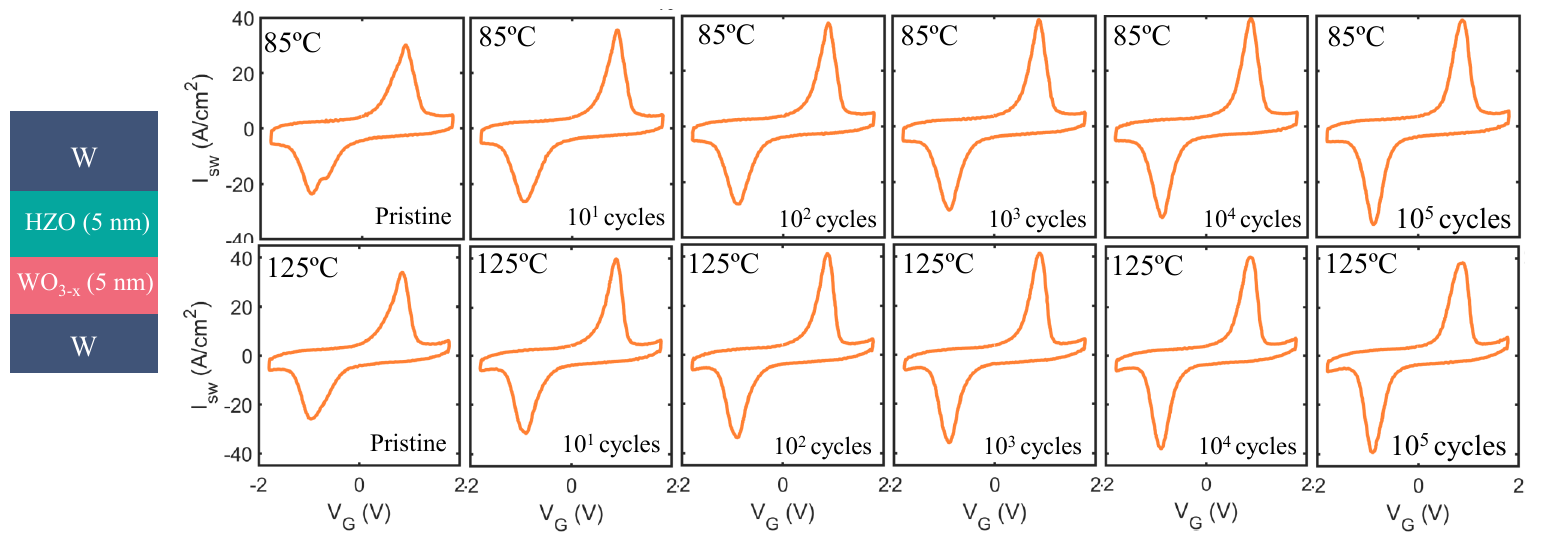}
\caption{I\textsubscript{SW}-V characteristics at 85 and 125\textdegree C from pristine to 10\textsuperscript{5} cycles for ALD deposited 5nm WO\textsubscript{3-x} device.   }
\label{IV_5nm}
\end{figure} 

Figure S10 shows P-V characteristics with cycling for 5nm ALD based WO\textsubscript{3-x} device measured at 85 and 125\textdegree C respectively. It shows better ferroelectric nature right from the pristine state compared to reference device as shown in Figure 4 (g-h).

\begin{figure}[H]
\centering
\includegraphics[ width=6.5in]{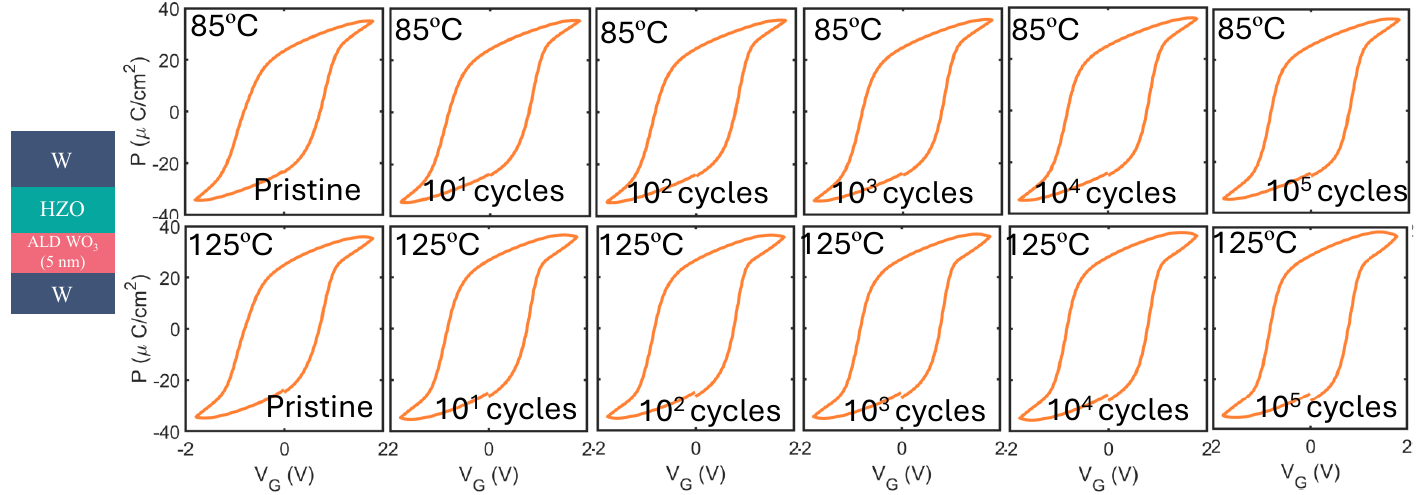}
\caption{P-V characteristics at 85 and 125\textdegree C from pristine to 10\textsuperscript{5} cycles for ALD 5nm WO\textsubscript{3-x} device.}   
\label{PV_5nmWO3}
\end{figure} 

When ten bipolar cycling pulses of ±1.8 V are applied to the ALD based 6 nm WO\textsubscript{3–x} device at 85\textdegree C and 125\textdegree C, the pure ferroelectric switching established at elevated temperatures is well preserved upon cooling to room temperature, as shown in Figure S11. 

\begin{figure}[H]
\centering
\includegraphics[ width=6.5in]{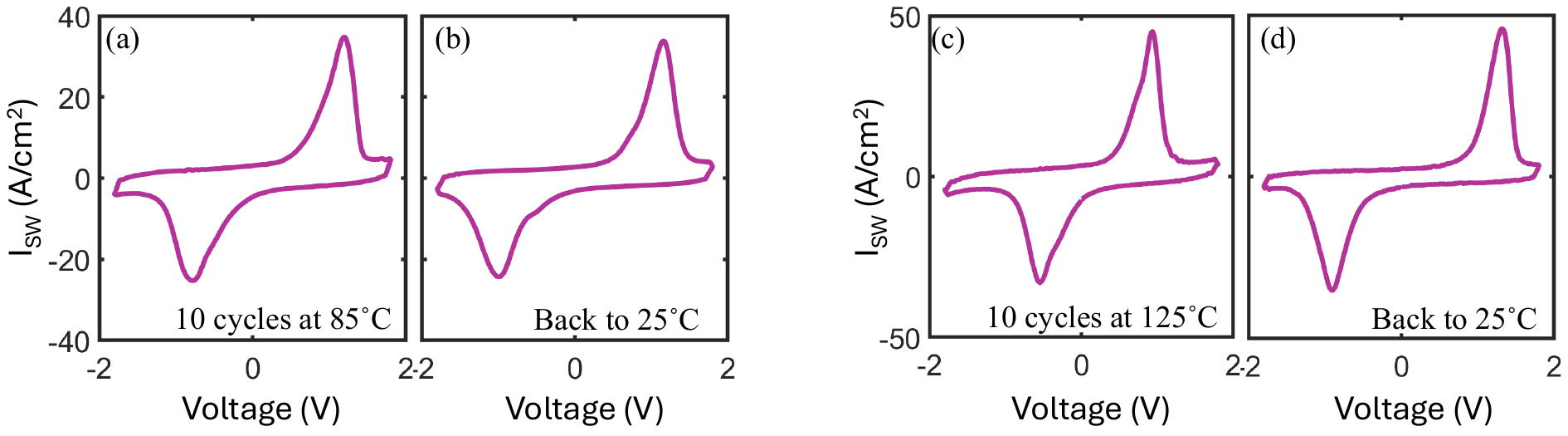}
\caption{I\textsubscript{SW} versus voltage characteristics of ALD 6nm WO\textsubscript{3-x} devices (a) after applying 10 cycles at 85\textdegree C and (b) bringing back to 25\textdegree C, (c) after applying 10 cycles at 125\textdegree C and (d) bringing back to 25\textdegree C.}
\label{HighT_backtoRT}
\end{figure}

I-V characteristics are measured across a temperature range of 25\textdegree C to 125\textdegree C up to 5x10\textsuperscript{6} cycles, using ±1.7V, 200 ns write pulse trains for reference and O\textsubscript{2} plasma devices, and ±1.8V, 200 ns pulse trains for ALD devices. The measurements from pristine devices are shown in Figures S12 (a-c). Figures S12 (d-f) show the fractional increase of current density at cycle number = n with respect to the pristine state (n = 0), $\Delta$J/Jo=(Jn-Jo)/Jo, Jn being the current density at 1V at n-th cycle, as a function of cycle number. The slope of $\Delta$J/Jo vs. cycles curves (which is a measure of trap generation rate with cycling)  is similar at room temperature for all the devices. However, the slope is much lower in all the WO\textsubscript{3-x} devices at elevated temperatures compared to the reference device. 

\begin{figure}[H]
\centering
\includegraphics[ width=6.5in]{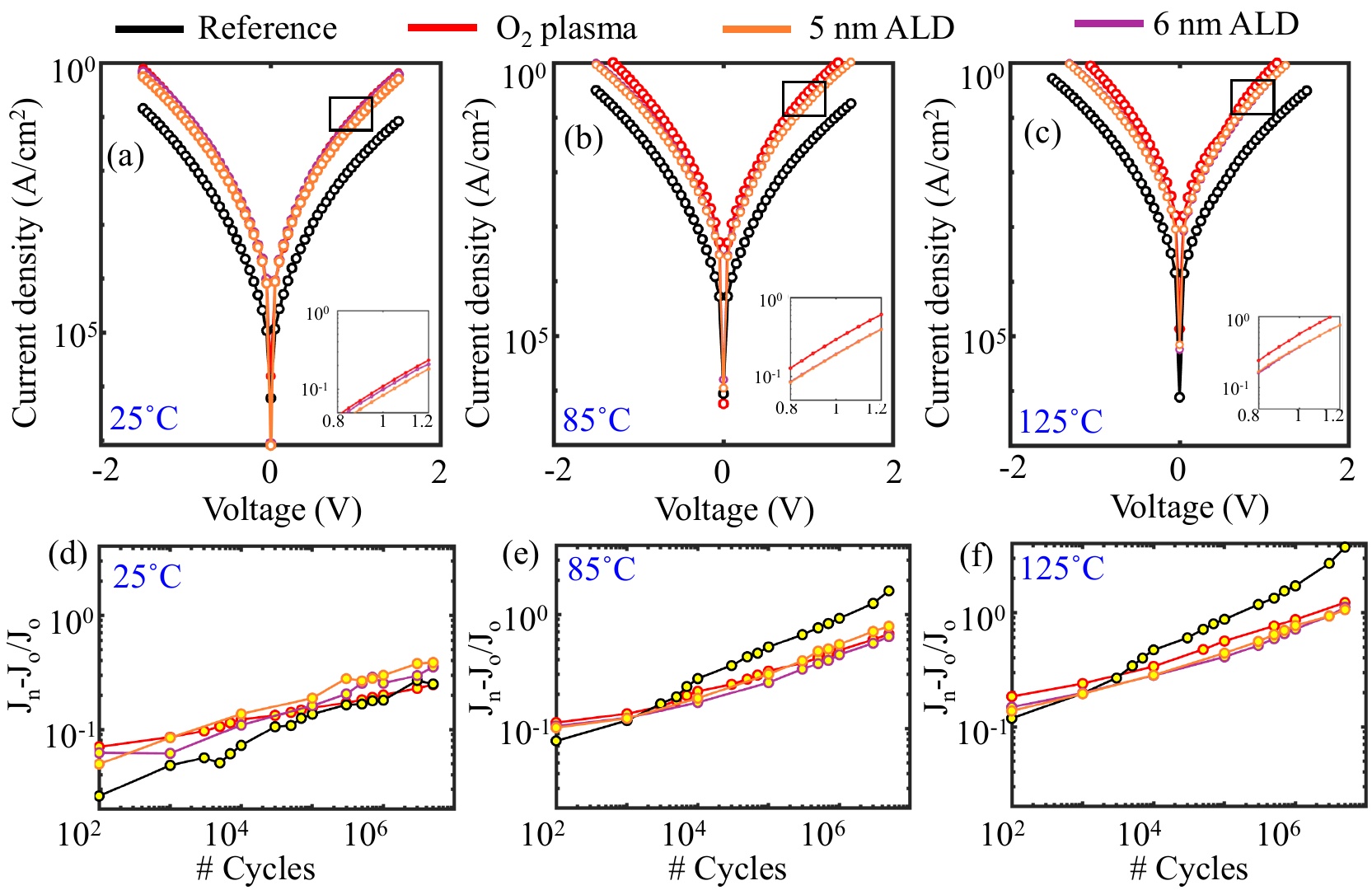}
\caption{(a-c) Leakage current density versus voltage at pristine state at different temperatures. The insets show the region labeled by the black box. (d-f) $\Delta$J/Jo with cycling at 1V at (b) 25\textdegree C (c) 85\textdegree C and (d) 125\textdegree C.}    
\label{Leakage}
\end{figure} 

Table S2 shows the energy barrier for t-to-m-  and t-to-o- phase transition in HZO and HZO strained to W and WO\textsubscript{3} lattices. Transitioning to m- phase is energetically costly than to o- phase from t- phase in all cases.

\begin{table}[h!]
\centering
\renewcommand{\arraystretch}{1.4}
\begin{tabular}{|c|c|c|}
\hline
 & \multicolumn{2}{c|}{\textbf{Phase transition energy (meV)}} \\
\cline{2-3}
 & \textbf{t-to-m phase} & \textbf{t-to-o phase} \\
\hline
HZO & 185 & 77 \\
\hline
HZO/WO$_3$ & 176 & 73 \\
\hline
HZO/W & 188 & 69 \\
\hline
\end{tabular}
\caption{Calculated energy barriers for tetragonal-to-monoclinic (t-to-m) and tetragonal-to-orthorhombic (t-to-o) phase transitions.}
\label{tab:phase-transition}
\end{table}

Figure S13(a) presents the calculated vibrational entropies for the HZO structures matched to the W and WO\textsubscript{3} lattices, respectively. The entropy of the W-matched system is consistently larger than that of the WO\textsubscript{3}-matched case over the entire temperature range. Figure S13(b) shows the corresponding phonon DOS.
As reported previously, the longer interatomic distances lead to softer vibrations, and the softening of phonon modes—manifested as a shift of the phonon DOS toward lower frequencies—results in higher vibrational entropy, since low-frequency vibrations increase the number of accessible vibrational states at finite temperature. Similar behavior in Figure 7a originates from the larger in-plane lattice matched to the W cubic structure under non-equibiaxial strain, which induces softening of the low-frequency phonon modes and consequently increases the vibrational entropy.

\begin{figure}[H]
\centering
\includegraphics[ width=6.5in]{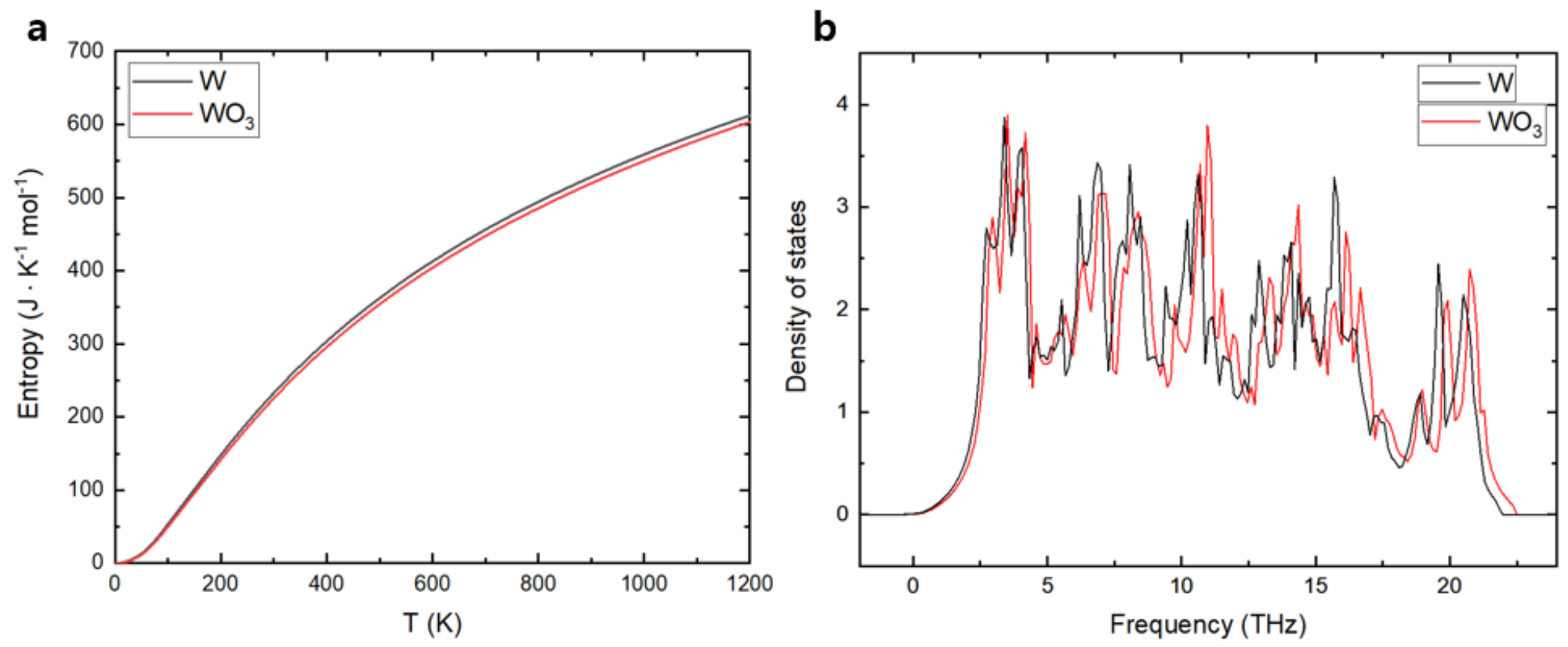}
\caption{(a) Vibrational entropies for the HZO structures matched to the W and WO\textsubscript{3} lattices. (b) Corresponding phonon density of states (DOS).}
\label{Entropy}
\end{figure}

In figure S14, it is clear that while transitioning from negative to positive polarization, the energy is higher at $\lambda$= 0 for HZO strained to W lattice compared to HZO strained to WO\textsubscript{3} lattice.

\begin{figure}[H]
\centering
\includegraphics[ width=6.5in]{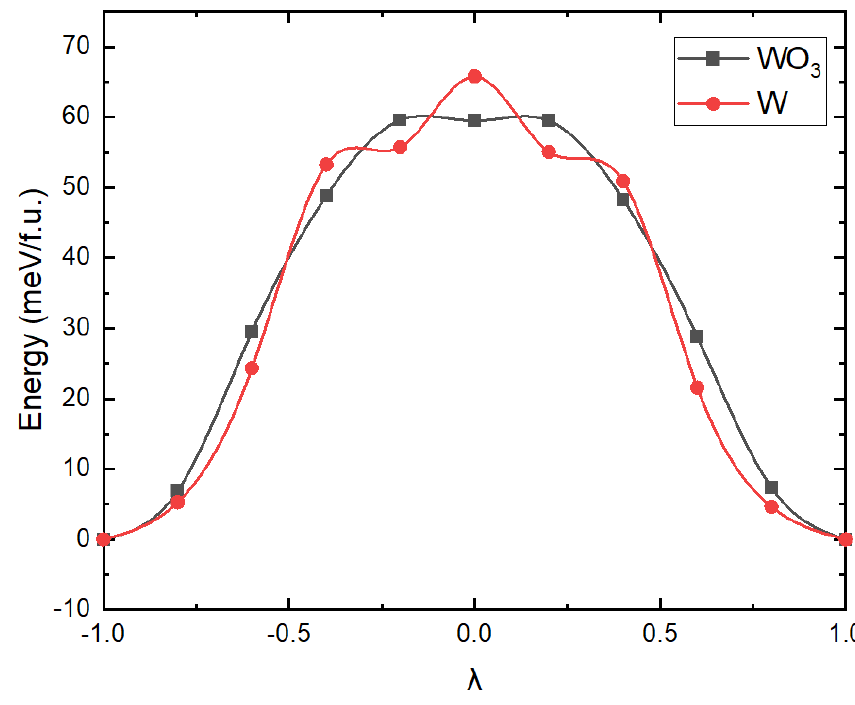}
\caption{Ferroelectric switching pathways for HZO strained to WO\textsubscript{3} and W lattices. $\lambda$ denotes the polar displacement, where $\lambda$ = $-1$ corresponds to the ferroelectric (negative) state, $\lambda$ = 0 to the paraelectric (zero) state, and $\lambda$ = 1 to the ferroelectric (positive) state.}
\label{energy_land}
\end{figure}



\end{document}